\documentclass[acmsmall]{acmart}
\AtBeginDocument{%
  \providecommand\BibTeX{{%
    \normalfont B\kern-0.5em{\scshape i\kern-0.25em b}\kern-0.8em\TeX}}}

\setcopyright{acmlicensed}
\acmJournal{PACMHCI}
\acmYear{2024} \acmVolume{8} \acmNumber{CSCW2} \acmArticle{477} \acmMonth{11}\acmDOI{10.1145/3687016}

\usepackage{xcolor}
\newcommand{\hi}[1]{\textcolor{black}{#1}}

\usepackage{caption}
\usepackage{subcaption}
\usepackage{graphicx}

\begin{document}

\title [Engaging Discontents in the Design of Family Meal Technologies]{"The struggle is a part of the experience”: Engaging Discontents in the Design of Family Meal Technologies}








\author{Yuxing Wu}
\email{ywu4@iu.edu}
\orcid{0000-0003-1149-6577}
\affiliation{%
  \institution{Indiana University Bloomington}
  \city{Bloomington}
  \state{IN}
  \country{USA}
}

\author{Andrew D Miller}
\email{andrewm@iu.edu}
\orcid{0000-0002-6152-6968}
\affiliation{%
  \institution{Indiana University Indianapolis}
  \city{Indianapolis}
  \state{IN}
  \country{USA}
}

\author{Chia-Fang Chung}
\email{cfchung@ucsc.edu}
\orcid{0000-0002-3374-2073}
\affiliation{%
  \institution{University of California, Santa Cruz}
  \city{Santa Cruz}
  \state{CA}
  \country{USA}
}

\author{Elizabeth Kaziunas}
\email{ekaziuna@iu.edu}
\orcid{0000-0003-0001-355X}
\affiliation{%
  \institution{Indiana University Bloomington}
  \city{Bloomington}
  \state{IN}
  \country{USA}
}

\renewcommand{\shortauthors}{Wu et al.}



\begin{abstract}
Meals are a central (and messy) part of family life. Previous design framings for mealtime technologies have focused on supporting dietary needs or \hi{social and} celebratory interactions at the dinner table; however, family meals involve the coordination of many activities and complicated family dynamics. In this paper, we report on findings from interviews and design sessions with 18 families \hi{from the Midwestern United States} (including both partners/parents and children) to uncover important family differences and tensions that arise around \hi{domestic} meal experiences. Drawing on feminist theory, we unpack the work of feeding a family as a form of care, drawing attention to the social and emotional complexity of family meals. Critically situating our data within current design narratives, we propose the sensitizing concepts of \textit{generative and systemic discontents} as a productive way towards troubling the design space of family-food interaction to contend with the struggles that are a part of everyday family meal experiences.
\end{abstract}





\maketitle

\section{Introduction}

Meals are a central experience of everyday life and represent a design space that has been conceptualized in HCI/CSCW research in diverse ways over the last 20+ years \cite{Wei2011FoodDinner, Wei2011CoDine:Dining, Barden2012TelematicPerformance, OHara2012FoodMeal, Ferdous2016CommensalityMealtime, Grimes2008CelebratoryHCI, Grimes2008EatWell:Community, Comber2013FoodHouseholds, Choi2010HCIEngagement, Farr-Wharton2014FoodWastage}. Understanding the highly situated nature of meal practices and re-imagining the ways people might cook and eat with technology draws together a wide range of design subfields, including domestic computing, sustainable HCI, and health informatics, to name a few. The \hi{design narratives} proposed in this rich literature often highlight different dimensions of human-food interaction. Weiser's vision of calm computing \cite{Weiser1991TheCentury}, for example, motivated many technologists to examine meal planning and preparation as a form of routinized domestic work that might be better automated through the use of computational artifacts and systems in "smart kitchens" \cite{Weber2023Designing, Mennicken2012Cooking, Sugiura2010CookingEnvironments, Angara2017Conversational}. 

Socially-oriented HCI/CSCW researchers, however, have demonstrated the \hi{varied} ways in which everyday domestic tasks (like making a pot of coffee to share \hi{with neighbors}) are significant to how people coordinate, communicate, and create meaning with one another in groups \cite{Crabtree2005MovingCSCW, Randall2021EthnographyEthnomethodology, Grinter2009TheNetworking}. Meals, then, are not just tedious forms of labor to be eliminated with technology but \hi{must be} understood as important social and cultural activities in their own right \hi{to be} augmented or supported through design. To this end, a number of HCI/CSCW studies over the past decade have sought to design meal technologies that support social and emotional interactions during people's meal experiences. An example of this literature can be seen in projects around supporting "commmensality," (defined as "the practice of sharing food and eating together in a social group such as a family" \cite{Ochs2006TheSocialization}) by facilitating novel mealtime conversations mediated by technology \cite{OHara2012FoodMeal, Barden2012TelematicPerformance, Ferdous2016TableTalk:Table} or virtually connecting people at mealtimes to enable shared moments of 'eating together' at a distance \cite{Grevet2012EatingCommensality, Judge2010SharingHome, Nawahdah2013VirtuallyDesign}.  

Influential in conceptualizing this wider turn to the social aspect in the design space of human-food interaction was Grimes and Harper's 2008 CHI paper \cite{Grimes2008CelebratoryHCI} in which they argued that many HCI/CSCW studies \hi{had} adopted a  "corrective" lens towards meal interventions (e.g., viewing technology as a way to address informational/work challenges like cooking inefficiencies or to fix problematic human behavior, such as unhealthy eating). While acknowledging the usefulness of a corrective framing for some food contexts, they noted its limitations and called \hi{on} designers to also create "celebratory technology" that promoted positive meal experiences. Grimes and Harper's conceptualization of celebratory technologies\cite{Grimes2008CelebratoryHCI} helped inspire new types of interactive technologies that centered values such as social togetherness, delight, creativity, and playfulness during meals\cite{Bertran2021TheExperiences,  Ferdous2017CelebratoryMealtimes}. \hi{It also demonstrates the continued importance of critically examining the assumptions of our dominant design spaces in order to challenge HCI/CSCW researchers and designers to explore diverse food behaviors and overlooked needs}.

While Grimes and Harper's reframing of human-food interaction \cite{Grimes2008CelebratoryHCI} offered a much-needed theoretical intervention, we would argue that important parts of the lived experience of meals are still not well accounted for in either the celebratory or corrective design framings. Through a study investigating the everyday food practices of families with children, we found many examples (aligning with Grimes and Harper's design ethos) in which mealtimes were experienced as enjoyable moments of family togetherness and fun. Just as often, however, family meals were emotionally stressful, tiresome, or even fraught encounters between children/parents, siblings, schools, workplaces, and grocery stores alike. And for all the utopian promises of smart kitchen technologies to reduce domestic work, for many of the parents in our study (particularly women), family meals remained exhausting forms of labor that \hi{left them feeling stressed and their personal needs unmet. In short, many of the meal experiences in households we spoke to were often self-described by families as \textit{a struggle}}.

The observation that it is difficult work to feed a family is nothing new; what is \hi{less} obvious is how best to address different types of domestic tensions (and even unhappy families) in system design. On one hand, struggle can be seen as a mundane and expected part of family life, especially during meals with children. Given that family members often have unique food preferences, differing nutritional goals, and (at times) conflicting emotional or physical needs, such everyday tensions are an important part of the context \hi{to consider} when creating technological interventions into mealtimes. Indeed, we would argue that in day-to-day family life, even commensality--the practice of eating together \cite{Ochs2006TheSocialization, Ferdous2016CommensalityMealtime}--is rarely a care-free celebration, but rather an ongoing negotiation of difference. On the other hand, there are other types of family meal struggles, such as food insecurity or gender-based inequities in domestic work, that while commonplace are \hi{also} deeply problematic. Such social and cultural issues require intervention but also suggest a different set of strategies and tactics than an overly prescriptive form of technosolutionism.

In practice, the work of feeding a family is messy and complex--and related domestic struggles can be both a natural part of family life and still stir up feelings ranging from mild resentment to trauma. However, in the creation of new food technologies, it can still be all too easy for designers to idealize family meals. \hi{\textbf{How might researchers and designers account for the types of food interactions that are not best framed as either a problem to be fixed or an experience to be celebrated? What conceptualizations might be useful in helping articulate new directions for the design space of family-food interactions and domestic HCI?}}

In this paper, we offer a two-fold contribution to the HCI/CSCW community. First, we report on empirical findings from an \hi{interpretivist} study of family meal practices that includes data collected from interviews and design sessions with 18 families living in the \hi{Midwestern region of the} United States.  Drawing on feminist scholarship that views \textit{feeding} as a distinct form of care work \hi{\cite{DeVault1991FeedingWork}}, our analysis calls attention to the social and emotional dimensions of family meals that have been understudied in current HCI/CSCW design approaches. Second, following in the footsteps of Grimes and Harper \cite{Grimes2008CelebratoryHCI}, we discuss the importance for designers to be able to grapple with a greater range of human-food experiences; and specifically, to address what we have come to think of as the troublesome (rather than negative) interactions found in family life. To this end, we propose two related sensitizing concepts to help the HCI/CSCW community create more socially nuanced and emotionally supportive domestic meal technologies: (1) \textit{generative discontents}, by which we refer to everyday interpersonal differences or tensions that might be amenable to negotiation or resolution within the family, and (2) \textit{systemic discontents} which involve those food-related struggles connected to broader societal and cultural factors that (for some) might necessitate interventions of resistance or change.

By focusing on troublesome experiences in family meals, our paper contributes to an emerging field of HCI/CSCW research concerned with understanding technology's emotional and social impacts on family life. So far, early research in this design space has focused on how technology shapes family relations between parent-child dyads or between parents or partners \cite{Derix2022FamilyRelationships, Yardi2011SocialUse, Hiniker2016NotRules}. Some scholars have noted, for instance, that technology can lead to negative social interactions at mealtimes, such as how children's use of digital technologies (like mobile phones) leads to disagreements between family members \cite{Hiniker2016NotRules,Chen2019UnderstandingTechnology}. This literature has generally taken a cautious stance on how designers might better integrate technology into the home, maintaining that adding more technology in contexts like family meals (even when carefully designed) is not always for the better \cite{Moser2016TechnologyMealtimes, Chen2019UnderstandingTechnology}. Indeed, there have been several calls for HCI/CSCW researchers and designers to adopt a "family-centered approach" that considers societal factors when designing technologies for the home \cite{Isola2012FamilyCommunities, Cagiltay2023FromDesign}. Our research is largely sympathetic towards family-centered research goals and design concerns. And to these conversations, we add a \textit{holistic} portrait of family meals requiring dynamic coordination and collaboration across many activities, including meal planning, food shopping, preparing, and cleaning. In this, our paper also offers HCI/CSCW scholars of families a critical reflection on some of the costs and consequences that come with intervening (even with good intentions) in the delicate relations of everyday domestic life.

\hi{Finally, it is important to acknowledge upfront that "family" as a form of social organization can be configured in any matter of ways \hi{such as} multi-generational, child-free, and chosen, etc. Given the limitations that come with any study, however, in this paper we focus our attention on understanding the particular needs of families that have school-age children in the home. As well, we also emphasize that feeding work is not only shaped by people's local geography or family size, but by complex sociocultural factors like race, class, religion, and education. Our paper tells a story of feeding that is predominately White, nuclear, and centered around the experiences of families living in a small university town in the United States. Not all of our participants fit that description, and we make a point to highlight their accounts; nevertheless, many more stories of feeding need to be told. We hope this paper helps facilitate that research by showing why multiple and nuanced accounts of diverse family life are both necessary and important in HCI/CSCW research and technology design.}

In the following sections of this paper, we first situate the reader within relevant literature on family meal technologies in HCI/CSCW, and then discuss the theoretical work that frames our understanding of family meals. Next, we provide an overview of our study design, including data collection and analysis. We then share our study findings, detailing types of disagreements, tensions, and conflicts around meals (as well as divergent visions for how to address these struggles through technology design). Finally, we discuss the usefulness of engaging discontents in technology design and hare possible design opportunities for supporting family meal experiences. We also reflect on the limits of technological interventions to resolve all care work tensions.

\section{Related Literature}

\subsection{Technology for family mealtimes}
\hi{The context of} family meals as an important time for shared social interaction has attracted much interest from the HCI \hi{/CSCW} communities. Researchers have studied various aspects of family mealtimes and explored \hi{design possibilities in using technology to augment} mealtime experiences, including facilitating content displaying and sharing during eating \cite{Ferdous2016TableTalk:Table, Ferdous2017CelebratoryMealtimes, OHara2012FoodMeal, Barden2012TelematicPerformance}, promoting healthy eating behavior \cite{Jo2020MAMAS:Recognition, Lukoff2018TableChat:Eating, Schaefbauer2015SnackFamilies, Grimes2008EatWell:Community, Randall2018Health-e-Eater:Families}, and supporting remote dining experiences \cite{Grevet2012EatingCommensality, Judge2010SharingHome, Nawahdah2013VirtuallyDesign}. \hi{Such designs demonstrate the} potential for technology to help with health needs and social interactions around family mealtimes, but also present specific challenges and limitations. For instance, \hi{HCI/CSCW scholars have pointed to the} challenge of acceptance and long-term engagement with existing mealtime technologies \cite{Schaefbauer2015SnackFamilies, TenBhomer20104Photos:Experience}, lack of understanding of diverse motivations and dynamics in shared family interaction \cite{Ferdous2017CelebratoryMealtimes, Lukoff2018TableChat:Eating}, and defining the scope of a "family meal" \cite{Ochs2006TheSocialization, Ferdous2016CommensalityMealtime}. \hi{Our study is in part motivated to better understand these limitations by unpacking the richness of family meal experiences.} 

\subsubsection{Defining family mealtimes} 
Shared eating practices serve critical social functions in everyday family lives and have attracted the most interest in HCI\hi{/CSCW} literature \cite{Sobal2003CommensalStudy, Fischler2011CommensalityCulture:}. The concept of "commensality", which is defined as “the practice of sharing food and eating together in a social group such as a family” \hi{has been a central idea in HCI/CSCW meal research} \cite{Ochs2006TheSocialization}. Many meal-related technologies focus on commensality to create new forms of social interactions during the consumption of a meal, both for facilitating in-person meals and for enabling remote dining experiences \cite{Barden2012TelematicPerformance, Nawahdah2013VirtuallyDesign, Ferdous2015PairingHousehold, Wu2020-MealChat}. In particular, Ferdous et al. applied the concept of commensality to examine family mealtimes to emphasize how commensality extends beyond the meal itself and explore new forms of supporting social meal practices \cite{Ferdous2015TechnologyOrdinary, Ferdous2015PairingHousehold, Ferdous2016CommensalityMealtime, Ferdous2017CelebratoryMealtimes}. This prior research describes the social needs and challenges that exist in shared dining practices and ways that systems might support them during the process, but can overlook important parts of a family meal (e.g., planning and preparation). To understand and design for a holistic experience of family meals, we \hi{extend our analytic view beyond} the concept of commensality \hi{to account for a wider range of activities and highlight the complex forms of coordination and} labor surrounding a family meal. In this paper, then, we use "family meal processes" to \hi{collectively} refer to the planning, purchasing, preparing dining, and cleaning activities that comprise a family meal. \hi{By more fully situating family meals back into the routines of everyday domestic life, we aim to provide a deeper understanding of how commensality is connected with other meal-related activities and forms of work (and why those relationships matter in design).}

\subsubsection{Celebratory technology}
Grimes and Harper \cite{Grimes2008CelebratoryHCI} proposed "celebratory technology" that emphasizes people's positive interactions and experiences with food through designing technology \hi{around values like} creativity, relaxation, and nostalgia. Motivated by this call, researchers have created systems inspired by the social and celebratory aspects of food. For instance, Ferdous et al. \cite{Ferdous2016TableTalk:Table, Ferdous2017CelebratoryMealtimes} designed TableTalk and Chorus that support sharing personal devices and forming communal displays on the dining table to stimulate conversation. \hi{Grimes and Harper also importantly note} several challenges in designing celebratory technology. Namely, when and how to introduce technology, how to account for the family dynamics, and how to measure the success of celebratory technology \cite{Grimes2008CelebratoryHCI}. \hi{In our study, we originally sought to} explore celebratory practices -- in particular, we set out to understand family members' expectations and goals for what they deem as desired meal practices and how they wanted technology to \hi{be a part of their daily meal experiences}. However, as our study findings will show, \hi{domestic tensions and disagreements around family meals led us to consider the ways celebratory approaches may miss important parts of the wider design story around family-food interactions.} 

\subsubsection{Attitudes towards mealtime technology use}
Despite the potential benefits that technology may add to mealtimes, parents and children may hold diverse attitudes toward mealtime technology use, which can lead to tensions within the family \cite{Chen2019UnderstandingTechnology, Ferdous2015TechnologyOrdinary, Hiniker2016NotRules, Moser2016TechnologyMealtimes}. A growing body of work \hi{in HCI/CSCW} has investigated how parents manage their families' use of technology, specifically in the meal context. For example, Moser et al. \cite{Moser2016TechnologyMealtimes} conducted a survey to understand attitudes about mobile phone use during shared mealtimes and identified various elements that influence family members' views about technology use at mealtimes. Hiniker et al. \cite{Hiniker2016NotRules} studied how technology rules are established in the home and discussed how tensions between parents’ and children's attitudes around technology usage suggest a need for finer control around contextual constraints for family technology practices. Chen et al. \cite{Chen2019UnderstandingTechnology} highlighted parental concerns that pervasive smart technology could displace parenting relationships and disrupt personal interaction between family members. Taken together, this literature emphasizes the negative impacts that technology adoption and usage can have on family life and calls for careful consideration in designing mealtime technologies that attend to the existing relationships and \hi{dynamics} in families. \hi{Our paper contributes to these ongoing discussions} by exploring design implications for family-food interactions beyond the dinner table, \hi{such as how meals connect to the wider contexts of work, school, and community. As well, our paper specifically engages children's perspectives on mealtime, shedding light on the views and goals children hold in relation to their parents' perspectives. In paying special attention to understanding children's experiences and attitudes (which often tend to be overlooked), we aim to provide both empirical and conceptual insights into creating supportive and engaging meal experiences for all family members.}

\subsection{Family conflicts around the use and design of domestic technology}

A family is made up of individuals, and an important part of navigating domestic life as a family means dealing with conflicts and reaching common ground as a group. Previous HCI/CSCW literature has investigated the differing perspectives within families, particularly around technology use as a source of disagreements and tensions \cite{Derix2022FamilyRelationships, Hiniker2016ScreenPreschoolers, Blackwell2016ManagingTeens}. Prior work, \hi{for example,} showed that parents and children could disagree on various topics, including how much, when, and what types of technology use are allowed, which can lead to frustration and \hi{numerous other} challenges in parenting \cite{Blackwell2016ManagingTeens, Vandewater2005NoYouUse, Mazmanian2017OkayAge, Beneteau2020ParentingDynamics, Yardi2011SocialUse}. 

As researchers look into how conflicts arise in the mundane everyday life of families, studies revealed the significance of understanding unique individual experiences that are embedded in the complex, rich dynamics of individual families \cite{Mazmanian2017OkayAge}. This literature calls for HCI/CSCW research that examines the family as a whole and accounts for diverse family dynamics. However, \hi{current efforts in this area have} primarily focused on parent-child dyads, which tends to overlook experiences of other parts of the family \hi{life (such as spouse/partner dynamics or sibling interactions)}. Motivated by this, recent work from Detrix et al. \cite{ChinDerix2020TacticsUse, Derix2022FamilyRelationships} focused on studying conflicts between parents who attempt to manage their children's technology use. However, little is known about how conflicts may arise for \textit{the entire family}, including the individual perspectives of each parent\hi{/partner} and the children. Our current study attempts to address this gap by taking a broader approach to understanding individual perspectives between all members of the collocated family. 

Additionally, we see an opportunity to uncover family dynamics beyond conflicts in current technology use; namely, how conflicts and tensions may affect the expectation and design of new/emerging technologies, \textit{before} they are introduced and adapted in the wild. In their investigation of technology rules in families, Mazmanian et al. \cite{Mazmanian2017OkayAge} highlighted the need for domestic HCI/CSCW research to dive into the messiness of family life, the power dynamics between individual family members, and the differing expectations and values for experiences. In this paper, we seek to proactively probe and understand the unique experiences, needs, expectations, challenges, and attitudes of individual family members that may need to be balanced with those of other members. We follow the call of Isola and Fails \cite{Isola2012FamilyCommunities, AlanFails2012TechnologyFamily} for more holistic approaches and ways to design for the whole family experiences with technology and adopting a "family-centered" \cite{Cagiltay2023FromDesign} approach in our work. \hi{In highlighting} family conflicts and tensions, this paper aims to provide a deeper understanding of how family dynamics are presented in current daily practices around a common family activity -- family meals. \hi{Furthermore, by} situating the study in mundane family meal practices, we provide a low-risk environment for parents and children to share \hi{a more honest and unfiltered} perspective on \hi{behavior and experiences that might not otherwise be openly discussed or shared (e.g. family fights).} We closely examine these otherwise socially undesirable \hi{moments} within families to develop design implications on what family tensions and differences mean for the design of future meal technologies \hi{and the field of domestic HCI.} 

\section{Theoretical Framework} \label{Theoretical Framework}

A growing body of HCI/CSCW literature has been interested in exploring the various practices of care work and the significance of care in design \cite{Toombs2018SociotechnicalCare, Karusala2021ThePractice}. Many studies have engaged with feminist theorists from  Science, Technology, and Society (STS) and ethics (such as Annemarie Mol, María Puig de la Bellacasa, and Joan Tronto) to apply a theoretical lens of care across a diverse range of design contexts, including examining the precarity of care work in marginalized health contexts \cite{Kaziunas2019PrecariousCare} and provide a critical perspective on patient experiences \cite{Burgess2022CareDesign}; care activities in Makerspaces \cite{Toombs2015TheMaking}, care as a part of civic engagement \cite{Meng2019CollaborativeDemocracy}, data science \hi{work practices} \cite{Wolf2019EvaluatingPractices, Wolf2019ConceptualizingDevelopers}, and repairing broken infrastructural relations \cite{Dye2019IfStreetNet}. 

In this paper, we draw upon the work of sociologist and feminist scholar, Marjorie DeVault, for a theoretical framing \hi{to examine} family meals as a site for system design. In her work, \textit{Feeding the Family: The Social Organization of Caring as Gendered Work}, DeVault argues for re-examining the (often trivialized and invisible) forms of labor that make up domestic life to acknowledge the skill, craft, and symbolic power of “the activities of everyday maintenance” in the home \cite{DeVault1991FeedingWork}. Particularly relevant to our analysis here is DeVault’s conceptualization of the family meal as \hi{an act of} \textit{feeding}, an \hi{everyday practice} which she argues is: (1) a complex and distinct form of care work; (2) inextricably embedded in family relations and interpersonal ties; and (3) often \hi{leading to} domestic tensions and/or conflicts when established routines are challenged or changed within a family \cite{DeVault1991FeedingWork}.

\subsection{Feeding as Care Work}

In her study of U.S. families\cite{DeVault1991FeedingWork}, DeVault found that shared meals in the home involve a number of different types of interactional activities. These include the repetitive/mechanical tasks typically associated with cooking or cleaning up, such as chopping vegetables or washing dishes, but also the necessary "thought work," required for managing routine food-related responsibilities, such as planning shopping lists and coordinating schedules. Taken together, these interactional activities come together in the work of feeding a family. DeVault explains that the act of "feeding," is inextricably connected to a person's social and material relations:
\begin{quote}
"Family meals are prepared for particular people. While \hi{`cooking'} is work that can be done alone, \hi{`feeding,'} implies a relatedness, a sense of connection with others. Producing meals is about serving family members in a double sense: the food provided for a family cannot be just any food, but must be food that will satisfy them" \cite{DeVault1991FeedingWork}.
\end{quote}

In this way, feeding should be seen as highly situated care work. For those individuals who take up this responsibility, feeding often involves cultivating a sensitivity toward meeting the different needs of each family member. For instance, the practices of feeding include making improvisational meal adjustments based on a variety of concerns (e.g., access to types of food, time and logistics, personal taste and health issues, etc.). Importantly, in DeVault's study (as well as ours), such care work was often \hi{socially} messy and \hi{emotionally} complicated for those who performed that labor, indeed trying to satisfy others through feeding often brought feelings of personal gratification, but also daily struggle and sacrifice.

\subsection{Feeding and Family Relations}

Central to DeVault's theoretical stance is the understanding of family as a form of social organization that cannot be extricated from the activities and daily work that produce it. She maintains: 
\begin{quote}
    \hi{"}As experience, \hi{`}family\hi{'} refers to the activities of daily lives; to small groups of women, men, and children who usually share material resources, who many work and play together, who sometimes love each other and sometimes fight. Experientially, households are sites of material interdependence, of care (both material and emotional), and often, of affection and respect. They are also arenas for social conflict, where power relations are reproduced. They are dangerous for some members, and anger and frustration are taught along with love\hi{"}\cite{DeVault1991FeedingWork}.
\end{quote}
The family meal, then, is not just a pragmatic matter of putting food on the table that meets everyone’s basic needs or being able to make everyone happy during breakfast, lunch, and dinner, but the enactment of being a parent, partner, or child. Seeing how the family meal is intimately embedded within family relations brings attention to the complexity of the act of feeding as care work, and how it can simultaneously be both an act of creativity and joy, as well as a source of tension and distress. 

\subsection{Idealization of the Family Meal}

Importantly, for this paper, DeVault's concept of feeding \hi{as care work} gives us a lens to critically examine the idealization of family meals in system design. While DeVault notes that \hi{"}overt conflicts\hi{" (e.g., sustained, explicit conflict about who would do the work or how it should be done)} about feeding are rare, her research did find that struggle is a common experience during family meals. Recounting a conversation she had with a single parent who was newly responsible for feeding his three children, DeVault described how this father shared his feelings of "overwhelming anxiety" that accompanied mealtime in his home. Another parent, she wrote, "told her about \hi
{`}the STRUGGLE involved in this feeding work,\hi{'} and the fact that “behind closed doors, dinner is often a nightmare" \cite{DeVault1991FeedingWork}.

Such daily struggles are not often recognized, argues DeVault, due in large part to the gendered nature of feeding, which is traditionally seen as a natural form of maternal affection rather than significant craft and knowledge work \hi{requiring expertise and skill}. DeVault maintains: "When conflict about housework does arise, it can be quite painful, at least partly because it carries so much emotional significance. Women who resist doing all of the work [...]risk the charge—not only from others, but their own minds as well—that they do not care about the family." While social norms and family demographics have diversified since DeVault's study was conducted and published in the early 1990s, recent studies of U.S. family life have shown the responsibility for domestic work is still primarily taken on by women, even during the COVID-19 pandemic \cite{Parker2021HowINQUIRIES}. For this paper, understanding feeding as gendered work highlights the potential emotional turmoil and social risk involved for \hi{particular groups} that comes with challenging societal expectations around family meals, or simply changing well-established routines in the home.

In the following section, we describe our study design for investigating the feeding work of families with \hi{school-age} children who are living in the \hi{Midwestern region of the} United States, and how technology might be \hi{designed} to reshape their family meal experiences.

\section{Family Meal Technology Project: Study Design and Methods}

This paper describes empirical findings from a research project exploring how to design technologies to support family meals with 18 families (n=61 individual participants). The goal of our investigation was \hi{two-fold}: (1) to document everyday activities and current practices involved in feeding a family, with a focus on understanding different perspectives and needs among family members in the same household; and (2) to explore family imaginaries of future meal technologies as a means of helping articulate desires and unmet needs. 

The data reported in this paper was collected during the early months of the COVID-19 pandemic. Given the necessity of social distancing, several members of our \hi{research} team (specifically the first, third, and fourth author) designed a protocol that allowed for a series of remote research activities with families. The sequence of the family meal activity, interviews, and design sessions described below helped us remotely build rapport with families and encouraged participants to reflect on their family meal experiences. We found it also allowed children to feel more comfortable elaborating on thoughts and ideas with their parents as well as the researchers \cite{Yip2016TheCo-design, Yip2017ExaminingDesign}. In the following section, we detail our data collection methods and analysis process.

\subsection{Data Collection}

\subsubsection{Participant Recruitment}
We sought to recruit a wide range of families by sharing our study on social media and through word-of-mouth in our local community. Our protocol targeted collocated families (i.e., parent(s) and child(ren) living in the same household) in the United States with at least one child between the ages of 9 and 14. The age range from middle childhood to early teens was intentionally selected as this period is an important time for children to develop independence and gain responsibilities in the home \cite{HoyA.W.Perry2012ChildDevelopment, YoungCDC, MiddleCDC}. Unlike toddlers or adult children, pre-teens and young teens are also more likely to regularly engage in a variety of meal activities in the home (beyond just food consumption), and be able to share their personal experiences and expectations in research activities that require sustained attention, such as group interviews and design sessions. No additional exclusion criteria were used during recruitment to allow for a diverse participant population. 

Families who expressed interest in our study were invited to join a remote introduction session with all family members present to learn about the study structure and expectations. We received verbal consent from families over Zoom\hi{\footnote{Online platform for virtual meetings: https://zoom.us/}} sessions to record these meetings, and each family member was given the option to individually decide whether they wanted to participate further in our study at the end of the introduction meeting. The study protocol was approved by our university institutional review board. Each eligible participant in the family (participating children and adult guardians) received a \$10 Amazon gift card per hour as compensation for their time in completing the interview and design session. 

\subsubsection{Family Demographics}
Participants in our study consisted of 18 families of different sizes, ethnicities, and education levels. While most (17 out of 18) families represented traditional nuclear families \cite{Bryant2016NuclearFamilies}, participants also included several diverse family experiences, including households with a single parent family (F13), mix-raced family (F8), and immigrant family (F9). The racial demographics of study participants were predominately White (14 out of 18), middle or upper-middle class (11 out of 18) families living in the Midwest region of the United States (16 out of 18). While the study participants generally reflect the local demographics of the community where the study took place--a small university town in a rural \hi{Midwestern} state--we fully recognize this limitation, its connection to broader challenges of recruiting during the pandemic, and its potential impact on our findings (further discussed in 4.3). 

\subsubsection{Family Meal Activity and Interviews}
Prior to their interview session, each family was asked to document the process of creating a family meal together, including planning, purchasing, cooking, eating, and cleaning as a household. Families were encouraged to take photos or short video clips over the entire course of their meal to capture details of their process and various food practices they felt were important. \hi{Families were given full discretion on which media material to include and share with the research team from this singular meal activity.} We requested families try to carry out this activity as similarly to their regular meal routines as possible; so while all family members were encouraged to participate, what this looked like was left to the discretion of individual families. Upon completing this meal activity, we asked each family member to reflect on their experiences and note what they liked and disliked about the activity. 

The activity reflections and photos (or video clips) were shared with the research team prior to the interviews. \hi{Families were encouraged to decide as a group on which photos or video clips to share. While we do not know who exactly made the decision, we try to overcome this by engaging the whole family to reflect on these data together during each interview.} Given that we could not conduct any direct observations of home life during the pandemic, these multimedia data allowed the research team to get a better sense of each family's current meal practices and grounded our subsequent interview questions as we could follow up on particular meal experiences, attitudes, and expectations toward eating together. The activity was also designed to provide all family members a chance to experience the processes as a group and elicit thoughts and reflections from everyone in the interviews afterward.

Within a week of completing the meal activity, the research team scheduled semi-structured interviews via Zoom with each family that lasted between 60-90 minutes. Each interview was conducted with the participation of the entire family (including parents and children) as a group to allow discussions to unfold and family members to each weigh in with their perspectives. All interviews (conducted by the first author) started by asking, “How did the family meal activity go?” and then followed up with questions that prompted the family members to reflect on how they felt about their current family meal experiences. \hi{To ease into the conversation and reflection, we reviewed the multimedia data with the family together using the screen share function of Zoom.} Recognizing that the collaborative meal activity may also have altered the routine or "normal" state of affairs in each family, we also asked each family member to reflect on how different the meal activity was compared to a typical meal experience at home. We found that the meal activity coupled with the interview successfully allowed all of the participants (especially children) to share details about their current meal routines and roles, as well as reflect on individual challenges and goals for future family meals. 

\subsubsection{Design Sessions}
Within a week of the interview, we conducted a follow-up design session (approximately 40-60 minutes) with each family where they were invited to imagine aspects of their current family meal process that they would like future technology to help with. Inspired by lessons learned from prior research on co-designing with families and children \cite{Druin1999CooperativeChildren, Yip2017ExaminingDesign}, we took a participatory approach in designing the session to encourage active engagement and discussion from all family members. This was a concern since we had to adapt design activities to the online format and wanted to be sensitive to both family power dynamics as well as children's attention span. We kept the session short with playful, open-ended prompts to engage all family members with the flow of activities. 

We asked family members to think about ways they would like technology to support their current meal process, as well as what an ideal family meal would look like with the technology they had in mind. We emphasized that “technology” can include out-of-the-box ideas (e.g., things that have not necessarily been invented yet) or may not even seem possible. Participants were given 20 minutes to sketch out their ideas individually and then share and discuss their sketches in a group for another 20 minutes. The discussion started with each family member describing their ideas in a "show and tell" manner, followed by Q\&A and discussion on how they felt about each design idea. The design sessions allowed families to make visible important (but often difficult to articulate) dimensions of family meals and helped researchers to better understand tensions between family member's different needs and desires.

\subsection{Data Analysis}
The remote interviews and design sessions were recorded and transcribed. During the transcription process, each participant was anonymized with IDs and pseudonyms (Table 1) that assigned a family name, first name, and participant number. For example, F1 is the Johnson Family of four -- mother Emily, father Daniel, and two children Lily and Ethan. To help form a holistic narrative for each family and their social ties, we used pseudonyms that incorporate meaningful demographic data from our participants and their relationship to each other within a family (such as age, gender, ethnicity, etc.)

Our analysis followed interpretive approaches in that our data collection and analysis mutually informed one another, and data was analyzed iteratively through an open coding process to identify significant concepts and themes \cite{Hsieh2005ThreeAnalysis, Braun2006UsingPsychology, hancock2001introduction}. In the first round, three members of the research team (specifically the first, third, and fourth authors) \hi{individually} coded transcripts from the first ten families to determine initial codes and cluster these into themes. \hi{Data were first open-coded and moved to a more inductive approach to form themes guided by the broad topics around family dynamics that researchers are interested in.} These initial team members then reviewed and discussed study themes, adapting and refining coding as new interviews and design sessions were conducted and analyzed. This process was repeated for three months until our study reached saturation, and we found repeating themes in our data and codes. 

In a secondary round of data analysis \hi{led by the first and second author}, the final dataset consisting of transcripts from 18 family interviews and design sessions was \hi{re-analyzed} by \hi{the research team}. Similar to the first round, the team met to discuss and refine existing codes and themes and to identify new concepts, eventually coming to focus on differences, disagreements, and tensions around family meals. \hi{As is typical in interpretivist research \cite{Clarke-framework} where rounds of data analysis and theoretical engagement with literature inform one another, our team drew upon feminist scholarship (such as DeVault's concept of feeding \cite{DeVault1991FeedingWork}) to help shape our understanding of gendered labor and domestic work and to develop sensitizing concepts from our analysis. Sociologist Kathy Charmaz \cite{charmaz2003grounded} wrote that sensitizing concepts "offer ways of seeing, organizing, and understanding experience; they are embedded in our disciplinary emphases and perspectival proclivities. Although sensitizing concepts may deepen perception, they provide starting points for building analysis, not ending points for evading it. We may use sensitizing concepts only as points of departure from which to study the data" \cite{charmaz2003grounded, bowen2006grounded}. 
In the field of HCI/CSCW, sensitizing concepts have been useful in highlighting characteristics of a particular type of activity or work \cite{Kaziunas2015-transplant, Steinhardt2015-Anticipation}, sociotechnical relationship \cite{Jack2017-Infrastructure}, or form of interaction with data \cite{Muller2019-data, Kaziunas2017-datafication}. Our articulation of \textit{generative/systemic discontents} is therefore aimed at being a conceptual point of departure for designers likewise interested in adopting a critical theoretical lens to reexamine the design space of family food interactions and domestic HCI.}

\begin{table}
\centering
  \caption{Family demographic information  (N=18 families/61 participating individuals, * represents non-participant family members)}
  \label{tab:participants}
  \footnotesize
  \begin{tabular}[htbp]{p{1.5cm}  p{1.5cm} p{3.6cm} p{3.6cm} p{2.1cm}}
    \hline
    Family & Race/Ethnicity & Parent (Gender, Age) & Children (Gender, Age) & Household Income\\
    \hline
    F1-Johnson & White & Emily(Female,N/A); Daniel(Male, N/A)* & Lily(Female, 11); Ethan(Male, 8) & N/A \\
    F2-Spencer & White & Jane(Female, 40); Dave(Male, 43) & Olivia(Female, 14); Liam(Male, 12) & \$75,000-\$99,999 \\
    F3-Green & White & Alice(Female, 46); Mark(Male, 48) & Sofia(Female, 11); Diego(Male, 8) & \$100,000 - \$149,999\\
    F4-Roberts & White & Jessica(Female, 39); Ben(Male, 45)* & Lisa(Female, 14); Noah(Male, 10) & \$200,000 and over\\
    F5-White & White & Rachel(Female, 41); Brian(Male, 43) & Hannah(Female, 12); Owen(Male, 10) & \$150,0000 - \$199,999\\
    F6-Field & White & Morgan(Female, 43); Sam(Male, 40) & Alex(Non-Binary, 10) & \$50,000 - \$74,999 \\
    F7-Jones & White & Lee(Female, 43); Robin(Male, 47) & Kai(Male, 16); Taylor(Female, 14) & \$20,0000 and over\\
    F8-Smith & White; Black/African American & Diane(Female, 46); Thomas(Male, 50)& Sam(Male, 12); Mia(Female, 11) & \$200,000 and over \\
    F9-Chen & Asian/Pacific Islander & Jen(Female, 36); Steven(Male, 41) & Logan(Male, 11); Bryce (Male, 6)* & \$200,000 and over \\
    F10-Carter & White & Rebecca(Female, 39); Kevin(Male, 38) & Jack(Male, 15); Erin(Non-Binary, 11) & \$100,000-\$149,999 \\
    F11-Scott & White & Melissa(Female, 46); Matt(Male, N/A) & Ava(Female, 11) & \$75,000-\$99,999 \\
    F12-Jean & White & Emma(Female, 46); Claude(Male, 49) & Jess(Male, 10) & \$75,000-\$99,999 \\
    F13-Hughes & Black/African American & Victoria(Female, 39)  & Henry(Male, 16); Will(Male, 14) & \$39,000-\$49,999 \\
    F14-Young & Asian/Pacific Islander & Laura(Female, 46); Richard(Male, N/A) & Jackson(Male, 15);  Caleb(Male, 11) & Under \$15,000 \\
    F15-Booth & White & Joanne(Female, 36); Tom(Male, 42) & Ella(Female, 9); Chole(Female, 11) & \$100,000-\$149,999 \\
    F16-Ellis & White & Stephanie(Female, 42); Andy(Male, 40) & Harper(Female, 10) & \$75,000-\$99,999 \\
    F17-Harrison & White & Allison(Female, 46); William(Male, 46) & Aiden(Female, 14); Lucas(Male, 12) & \$50,000-\$74,999\\
    F18-Warren & White & Kelly(Female, 39); Jason(Male, 39) & Molly(Female, 12); Jacob(Male, 10)& \$75,000-\$99,999 \\
  \hline
  \end{tabular}
\end{table}

\subsection{Limitations}

Because we recruited participants via snowball sampling and location-based online platforms, such as Nextdoor, the majority (78\%) of our participants are \hi{White} families. Although this aligns with the \hi{W}hite-dominated demographic (75\%) of the Midwest region where the research took place, our participants do not represent the national demographics of the United States. Our participants were also skewed towards a higher household income. While families in our study (35\% \hi{of households earned} between \$50K-\$100K) align with the regional composition of middle-class families, these families may have better capabilities in accessing and providing meals \hi{than more under-resourced communities}. \hi{We recognize the limitations of our participant demographics. In this paper, we are mindful to acknowledge the importance of family experiences we didn't capture, while still seeing value in exploring the wide range of social and emotional perspectives on family meals that are represented in our dataset. Future HCI/CSCW work is needed to ensure diverse racial, cultural, and socioeconomic family experiences are represented in technology design.}

As well, most families in our study were what is typically considered traditional nuclear families as well as citizens of the United States. Although our participants included other types of family experiences, such as a single parent family (F13) and immigrant families (F9, F14), given the limited scale of the study and the wide definition of a "family" our findings can not substantially capture the mealtime experiences of these populations. Despite our sample’s limitations, our data surfaces some interesting observations on how diverse types of families manage family meal processes (i.e., F13, F9), and hope to inspire future research in exploring how family meal experiences might be different for various types of families (i.e., LGBTQ families, immigrant families, divorced families). 

\hi{Finally, due to a desire to capture a holistic portrait of families, we asked all family members to join the mealtime activity, online interview, and design sessions together.} All interviews and design sessions were conducted in groups where children and parents met with researchers in one video conferencing call. \hi{Although this allowed us to better document and understand different family member perspectives (especially from children and their siblings), this research expectation may have limited certain families’ availability to participate in our study. We also} acknowledge the potential influence of power dynamics between parents and children during the \hi{group} interviews. We \hi{sought to} address this \hi{limitation} by meeting with all members during introduction sessions prior to the interviews to build up trust and rapport. We also took care to probe members based on their reflection survey responses from the meal activity. \hi{Furthermore, multimedia data were collected from a singular, one-time meal activity which we asked the families to perform. Future work could look at more long-term engagement to identify any differences and similarities compared to our isolated design activity on family meals.}

\hi{Despite these limitations, our study aims to help nuance our understanding of designing for shared family meal experiences. By highlighting the types of domestic tensions and differences we found in our study, we hope to inspire future work that explores different sociocultural, economic, and environmental dynamics that may impact not only the future of family meal technologies but also how we understand and engage with complicated family dynamics and "struggles" in the design community.}

\section{Findings}

In this section, we share feeding stories that highlight the different (and often conflicting) ways parents/partners and children experience family meals. First, drawing on interview data, we detail how everyday disagreements arise amid the work of feeding a family--such as deciding what counts as a healthy meal, coordinating schedules for mealtimes, and determining how best to eat together as a "family." We also describe how the activities of feeding are embedded within family relations for our participants, noting the types of frustrations and concerns that can arise amid trying to change or challenge those established roles and responsibilities. Secondly, reporting on data from our design sessions, we share the ways families imagined technology might support (or possibly harm) the experience of family meals. Taken together, our findings reveal a nuanced picture of family meals that helps illustrate the social and emotional complexity of feeding as care work.

\subsection{Everyday Differences and Disagreements around Family Meals}
\subsubsection{Negotiating nutritional goals and being a "healthy" family}

While "healthy eating" was initially often presented as a shared family value or goal, the majority of our participants described struggling with ongoing tensions about what this meant for everyday meal preparation. Families, for instance, often disagreed about how to cultivate healthy eating practices as a household. Indeed, parents often expressed vastly different understandings of what counted as "healthy" or "unhealthy" food for their children and placed varying levels of importance on ensuring that the meal they made met a particular nutritional standard. 

The Spencer family's (F2) feeding story exemplified such tensions. As a couple in their forties and parents to two teenage children, Jane and Dave Spencer found themselves in disagreement about their family's commitment to healthy eating. During our interview, Jane spoke passionately about how much she valued eating vegetables and "as a mother" sought to make sure her children routinely had access to these foods as well. From her perspective, feeding a family meant making careful food selections that optimized nutrition:\textit{"I make it a huge point to make sure we never have a meal without vegetables. That's super important to me. But getting to that every single time is time-consuming."} 

Dave, on the other hand, took a more relaxed approach to family nutrition, seeing his children as resilient (even if they ate less vegetables than his wife wanted during meals). "\textit{I would say that my perspective on the kids eating healthy is way less important than} [Jane]," he admitted. He further explained his philosophy on feeding: "\textit{I think that healthy eating is way less important for children and much more important for older folks. Because it impacts our health more aggressively and dramatically than it does for kids. Kids are very versatile. They'll be fine. They can literally eat butter every single day for a year and they would be fine.}"

Jane admitted that Dave's lack of serious commitment to nutrition emotionally weighed on her. In response to his declaration of their children eating butter, she both laughed \textit{“I don't think that's true..."} and then lamented: \textit{"I have tremendous grief when } [Dave and the Spencer children] \textit{throw chicken nuggets and french fries, or tater tots in the air fryer. And I'm like, oh my gosh, you cannot do that again! Since they were little, I made sure that they had the proper nutrition and cultivated healthy eating."}

Similarly, many other families in our study noted that having divergent understandings of how strictly parents should enforce healthy eating habits for their children was a source of stress in their homes. At one point during a family interview, a young boy shared with his family that he would like to be able to have ice cream more often. We listened with curiosity as his parents debated about whether it was a good idea or not for their son to have ice cream after dinner for several minutes, only to end up still divided on the issue. While the boy's father thought \textit{"it wouldn't hurt"} to let their kids have ice cream, his mother was far from convinced. "\textit{But we're trying to teach him healthy snacking!"}, she exclaimed with frustration. \textit{"He won't be able to learn a habit if you keep indulging him"} (F12).

With parental views about what counted as nutritious food or how to best to cultivate good eating habits in their children often at odds, we found the goal of creating a shared family commitment to healthy eating was in constant negotiation. While a mundane part of everyday meal preparation, it was also striking how these seemingly minor domestic squabbles around meals held profound meaning. Indeed, for some of our participants, feeding unhealthy food to their family was seen as neglect (a profound act of \textit{non-care)} that left them feeling distressed, angry, and even grieving in their desire to be responsible parents and healthy role models. 

\subsubsection{Coordinating mealtimes and worrying about when to eat together}

Other everyday disagreements often emerged around the coordination of family mealtimes. When asked about what they most prioritized in family meals, parents in our study overwhelmingly spoke about the importance of eating their evening meal together. This goal, however, presented a complex organizational challenge as families sought to balance multiple schedules and competing commitments between work, school, extracurricular activities, and just spending time together. Among our participants, family members expressed that different expectations around family mealtimes were an ongoing source of tension in their relationships.

We found that one adult in the household typically assumed responsibility for managing the coordination of mealtimes, and that changes to mealtimes for them were a complicated logistical process that often led to stress and frustration. While many men cooked for their families, in our study, we found that this coordination role (as well as the task of preparing the main shared meals) was still most often taken on by women. Indeed, during group interviews when women shared about the complexity of coordinating family meals, many of their male partners confessed to being unaware that even relatively minor schedule alterations (such as getting home 15 minutes late) impacted the finely orchestrated timeline of tasks that had to be accomplished to put together a family meal.

For many of our participants, the event of the family meal was difficult to pull off each day even with good communication and a shared commitment. As a working mother with two school-age children, Joanne (from F15) explained to us that organizing mealtimes in her home is \textit{"a lot harder than you think and caveats can happen very often."} While she and her partner both do their best to keep to a schedule, plans often go awry:
\begin{quotation}
\textit{"Sometimes I'll find out we miss an ingredient for a dish, and I will ask my partner to get it on his way back since he drives by Kroger} [a grocery store], \textit{but he often comes back so late that I had to make something else, or had to serve dinner late. I expect everything to be on time, but timing is tricky."}
\end{quotation}

Participants who prepared family meals routinely found themselves needing to come up with creative meal substitutions (F6), directing other family members on last-minute shopping needs (F8, F15), and rearranging after-school schedules (F16). Failure to coordinate these complex and changing factors resulted in much delayed or even missed meals and the additional emotional labor of dealing with tired partners and hungry children (F11, F12). For those families in our study, eating separately--especially for the evening meal--was something participants sought to avoid, as routinely not eating together resulted in feelings of anger, disrespect, and frustration by more than one family member.

Importantly, we found that the coordination work (e.g., keeping to a schedule) needed to create a family meal often sat in tension with the wider goal of being together. An excerpt from our interview with the Green family (F3) -- partners and parents Alice and Mark -- captured this dynamic. Discussing how she navigates a busy day, Alice (who was the family's main meal preparer) confessed that making sure family dinner was always on time has become her "mission":
\begin{quotation}
\textit{"Time is 100\% a priority for me. I need to make sure} [Green family's two young children] \textit{are done with dinner and drop them off for the soccer club at 6 p.m. So I have exactly one hour to get everything ready. It's always crammed and I know they} [the rest of the family] \textit{don't care much about being a few minutes late, but I made it a mission for myself."}
\end{quotation}
In response to Alice's quest for punctuality, Mark reflected: \textit{"}[Alice] \textit{puts too much pressure on herself. I really don't think that's a big deal. So what if we finish the dinner late? I would rather make sure we enjoy the time to eat together than worry about the time all the time. I think the kids would understand."} Viewed through the lens of care, in this feeding story, Alice experienced the family meal as intertwined with other responsibilities and tasks, such as ensuring their children make it to practice on time. Care, then, requires "cramming" all these activities into a timeline where no one loses out. For Mark, however, who often got off late from work and was mainly in charge of cleaning up after eating, the family meal represented a meaningful moment of pause and respite from the pressures of a hectic schedule. The family meal, here, can be seen as a site of conflicting visions of care that generated (for each family member) a set of worries and concerns around coordinating mealtimes.

\subsubsection{Navigating commensality and non-idealized family interactions}
  
As a lot of HCI literature has previously explored, the act of eating together -- commensality -- is often seen as the core experience of the family meal \cite{Ferdous2016CommensalityMealtime, Bertran2021TheExperiences, Grevet2012EatingCommensality}. In our study, families also shared the sentiment that 'eating together' was not just a pragmatic part of their daily routine, but also an important ritual and time for connection, celebration, and enjoyment. In the course of interviews, however, it also became clear that idealized visions of sacred family time were routinely disrupted by the messiness of everyday life, including surly teenagers, exhausted parents, and bored partners. Different individual needs and desires often led to disappointment in (and occasionally frustration with) mealtime interactions.

During an interview with the Jones family (F7), for example, parents Lee and Robin both shared how they deeply valued the opportunity to check in on their children's lives during family meals. \textit{"I enjoy spending time with my family,"} explained Lee, \textit{"and hearing things about everyone's day, and their take on how the day has played out."} Robin strongly agreed, \textit{"For me, it's all about spending time with family...chatting. So the connection part is really important."} As parents, then, they found it a bit disheartening (although not surprising) that their teenage sons tried to eat their meals as quickly as possible. Fourteen-year-old Kai offered a blunt reflection on family meals: "\textit{I don’t really care for the conversation...at least not the check-ins we have now. I just want to eat food. That’s all}.” Taylor echoed his older brother's views noting that conversations at family meals often felt perfunctory and less than inspiring. \textit{"I would like to have something more exciting or fun for our family meal. So it's less boring and more engaging."}

Creating a shared vision for family meals often meant compromising on individual preferences about what made meals enjoyable. In the Field family (F6), for instance, partners Morgan and Sam each valued different aspects of a family meal. A self-described "food lover," Sam delighted in having novel food experiences and creating different dishes for her family to try out together. In contrast, Morgan shared that he would be happy eating hot dogs or peanut butter and jelly sandwiches every day as long as he could have a good conversation. He explained:
\begin{quote}
\textit{“I would argue that the thing that I find the most valuable about mealtime is not about the actual meal itself. The food is somewhat irrelevant. It could be the same thing, every single meal. It's about the people you talk to, the things you hear, and the things you learn. Those are more interesting to me. And so if you ask me, what makes a mealtime enjoyable? I would say by removing the other distractions and focusing on the core element of what I find interesting in mealtime, which is the conversation and interaction with other human beings -- my kid and my wife."}
\end{quote}

While Sam valued conversation as well, she found Morgan's vision of family meals to be less than inspiring. "\textit{That would be a very sad life for me,"} she maintained. "\textit{I have to have interesting food. I'm not okay with eating peanut butter and jelly sandwiches every day of my life. So I guess I'm the one who values the other aspects...The food comes before the conversation.}” In contrast to family meals as times of shared togetherness and unity, the Field's feeding story highlights the ongoing care work of attending to differences in people's everyday experiences of commensality. Being sensitive toward the individual needs and preferences of family members and navigating tensions are often invisible but routine parts of many family meal experiences. Sometimes, differences during family meals can be successfully negotiated (after all, good food and conversation are not mutually exclusive goals); in other situations (such as encouraging unenthusiastic teens to communicate), finding a path toward happy family dinners may require additional creativity, fortitude, and forms of support. 

The feeding stories we have discussed so far represent relatively innocuous domestic tensions and are very much a part of the lived experience of most families, especially those with children. However, they are also significant in that they illustrate how, within a family, feeding work is often a struggle. Family members held opposing views on healthy eating, worried about finding time to eat together, and had divergent priorities and needs about what they deemed important, meaningful, or enjoyable about eating together. In the next section, we turn to explore the way the work of feeding is embedded in particular family relations and the struggle involved in questioning or changing established roles and responsibilities.

\subsection{Confronting Roles and Responsibilities in Family Meals}

As noted previously in Section \ref{Theoretical Framework}, the family meal is far more than simply eating together: it requires a lot of mental effort and physical labor to make it happen each day, including meal planning, shopping, preparing, cleaning, etc.  As well, these tasks can have a number of social and emotional dimensions as this feeding work is embedded within intimate family relations and wider sociopolitical contexts. As our participants shared the ways they coordinated and collaborated to make meals, we saw how everyday food activities were tied to particular roles and responsibilities that could be both highly meaningful and (at times) unsatisfactory or problematic. Trying to change these established relations around family meals, however, was often a struggle.

\subsubsection{Gendered experiences of meal planning and preparation}

While families reported different ways of sharing the labor of meals (e.g., someone going food shopping while another is responsible for cleaning up), we found that it was typical for one person in each household to take on the primary responsibility of meal planning and preparation (a task that involved a great deal more time, organization, and physical effort). In our study, the main "meal preparers" in 15 out of 18 families were women, the majority of whom balanced this domestic labor not only with parental responsibilities but also full-time employment. While many of the women in our study spoke positively about the importance of family meals in relation to their roles of being a 'mother' or 'wife,' many found themselves confronting wider social and cultural norms that still problematically dismiss domestic labor as "women's work." Having one's skills and efforts undervalued by others rankled. 

Women who were responsible for meal planning and preparation--especially during a busy work week--often described being both physically and emotionally exhausted. The responsibilities of meal planning and preparation are particularly heavy when families lack access to the resources and support they need to routinely put food on the table. The feeding story of the Harrison family (F17) illustrates some of the emotional costs that come with doing this work amid personal challenges. Allison, a working mother of two teenagers in her late 40s, shared that while she generally tried to stay pragmatic about getting done everyday tasks like shopping and cooking, occasionally her circumstances--which include dealing with several health issues amid financial concerns--can make it difficult to cope. She explained: \textit{"Generally, I'm okay with the role. It needs to get done. But there are times when I work all day and get home and I don't want to cook. And }[my partner] \textit{and the kids will suggest us to do takeouts. It just makes me feel bad that we're spending more on unhealthy food like burgers."} The guilt and conflict Allison described in buying "takeouts" when she was feeling worn down can be seen as an internal struggle with societal expectations about what a "good mother" does to take care of their family, including feeding children a healthy, home-cooked meal.

Confronting societal pressures around motherhood was often challenging, particularly when women felt the labor in preparing family meals was invisible and undervalued, not just by society at large, but also within their own homes. The Chen family's (F9) feeding story illustrates the complex nature of feeding work as a valuable act of care that can also still diminish the person who performs that labor. As a family of first-generation Chinese immigrants to the United States, Jen shared how she found herself both appreciating and critically reflecting on their family's cultural food traditions. Having a demanding professional career while being a mother, for instance, she struggled with the commitment of cooking meals every day for her husband and two young sons:
\begin{quotation}
\textit{"Thinking about it, someone preparing a meal for someone else. It's a big commitment especially (if) it's on a daily basis, and it can easily get boring. Another part is when everything is washed up...people tend to forget about how hard you work for them...That is one thing when I was in college and my aunt was taking care of me. And she was telling me, ‘I have been a housewife for 20, 30 years. The thing is, no one appreciated it because when people finished eating and I washed all the dishes, the day was gone and people forgot about it. That’s true in my family now. I have to be committed and find time for family dinners but they can so easily forget about it. I love my family and I like cooking for them, but I want to make sure my effort is recognized. You know, teaching that to the kids as well."} 
\end{quotation}

While Jen's commitment to meal preparation was rooted in a love for her family and heritage, she was also determined to avoid the unfulfilling experiences of the older women in her family who had their domestic contributions so easily erased. Instead, she sought to teach her children (and remind her husband, Steven) to recognize and appreciate the labor involved in her feeding work. Going beyond acknowledgment to change these established family roles, however, has proved more challenging. During their interview, Steven was quick to note that it's "\textit{not fair}" for Jen to do all the family cooking, but also admitted that they both felt stuck in their current arrangement. "\textit{I wish I could help more}" he explained. \textit{"I'm trying, but it's hard... I have long work hours, and my cooking sucks. We also don't want to do takeouts every day, and the kids like} [Jen]\textit{'s dishes.}" Amid the daily pressures of raising two young children and managing demanding careers, finding a way to better share the responsibilities of feeding was a struggle this family did not feel able to take on (at least for now).

\subsubsection{Transitions in family roles and feeding responsibilities}

While one's partner's role in meal preparation was often treated as calcified due to a number of complex external factors like gender norms, work schedules, and skill sets, children were naturally seen to have changing roles and responsibilities, especially as they matured. Many parents in our study, for instance, expressed a strong desire for all their children to play a more active role in meal preparation. Learning how to cook was seen as both a way for kids to help out at home and an important skill for their future independence -- a necessary act of care.  

Rarely, however, were all members of the family on the same page with what these new roles and responsibilities should look like, and how feeding work was best taught. Some parents found the idea of their children let loose in their kitchens nerve-wracking, concerned about what meal tasks children could safely take on. There were also varying levels of interest and motivation expressed by children for doing this work.  

The feeding story of the Smith family (F8) exemplifies the uncertainty around how parents can help children take on new responsibilities in meal preparation. As a mother, Diane's motivation to start cooking lessons with her kids came from her own lack of cooking experience when she left home for college -- a situation that made it a struggle to take care of herself in a healthy way. \textit{"It's a crucial life skill to have,"} she emphasized to her pre-teen son and daughter. At twelve years old, her son Sam, however, was not particularly enthusiastic about starting cooking lessons: \textit{"I'm just not into cooking. Especially if it's just normal cooking, it sounds boring."} 11-year-old Mia, however, seemed to embrace the invitation into the kitchen. \textit{"I think it will be interesting,"} she admitted. "\textit{My classmates sometimes mention they get to pack their own lunchbox, and I kinda want to do that too...I wish there were more cooking for me to do, like learning how to use the grill."}  While wanting his children to take on more responsibility, as a concerned father, Thomas, was hesitant about letting either child take on tasks unsupervised and worried this would ultimately create more work for him and his wife. \textit{They're clumsy,}" he admitted with a smile, "\textit{and I don't really trust them in the kitchen alone}." 

In summary, within a family, each person plays a different role in the complex process of the family meal. While many families sought a fair division of feeding work, this ideal was often a struggle to achieve in practice. As family members described their individual roles, current meal preparation practices, and future goals, we found traditional gender expectations were often stubbornly embedded within mealtime activities and that this feeding work was costly for women both in terms of physical strain and emotional stress. Furthermore, even when changes were a much-desired outcome, family members had difficulties in transitioning away from established mealtime roles or in adopting new responsibilities, such as empowering children to take a more active and independent role in meal-related activities. Next, we turn to examining the ways families envision technology as a possible solution/intervention to their ongoing struggles with feeding work.

\subsection{Fraught Visions for Family Meal Technologies}

During design sessions conducted with families, we asked participants to imagine how technology might be used to make family meals better experiences for them. As children and parents envisioned future meals with sketches of digital games, virtual reality, or robot chefs, we found that the varying meal-related values, roles, and responsibilities of individual family members (as described in sections 5.1 and 5.2) represented very different starting points for conceptualizing the design of family meal technologies, as well as concerns about how that technology could potentially impact family life.

Participants who took on the role of primary meal preparer (in this study, mothers in all but three families) generally imagined technology could assist with the functional aspects of the meal, supporting their current informational and work activities of organizing food shopping, schedules, and recipes, as well as cooking and clean-up tasks. In contrast, the other partner (here, mainly fathers) often focused on how technology could support social interactions during mealtimes that emphasized emotional bonding and a shared sense of being together. The sketches of children explored novel ways of enhancing meals with technology, including a number of gamification concepts to make meals more fun and engaging.

We unpack several sketches from our design sessions to show how technological interventions into family meals represented fraught design visions in which participants both desired technologies to help alleviate daily frustrations and solve domestic problems, but also feared what those changes might mean for the well-being of the family. Taken together, these sketches captured social and emotional tensions important for understanding the family meal as a complex site of care and what it might mean to reconfigure those socio-material relations in new ways.

\subsubsection{Reconfiguring feeding work}
\begin{figure}
     \centering
     \begin{subfigure}[b]{0.48\textwidth}
         \centering
         \includegraphics[width=0.9\textwidth]{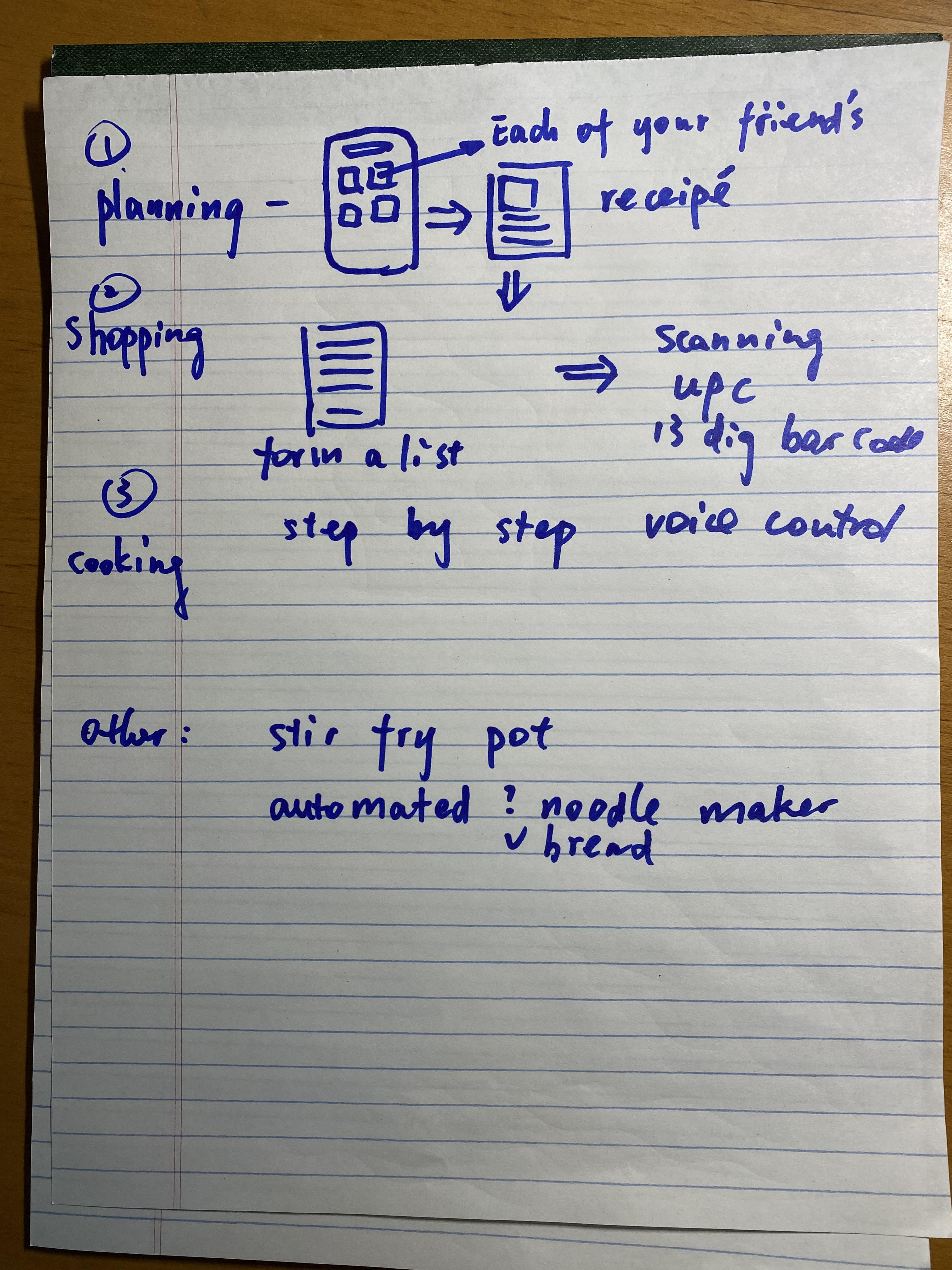}
         \caption{Jen's (mother from the Chen Family) design sketch, planning, shopping and cooking a meal with a smart kitchen}
         \label{fig1a:Jen's sketch}
     \end{subfigure}
     \hfill
     \begin{subfigure}[b]{0.48\textwidth}
         \centering
         \includegraphics[width=0.9\textwidth]{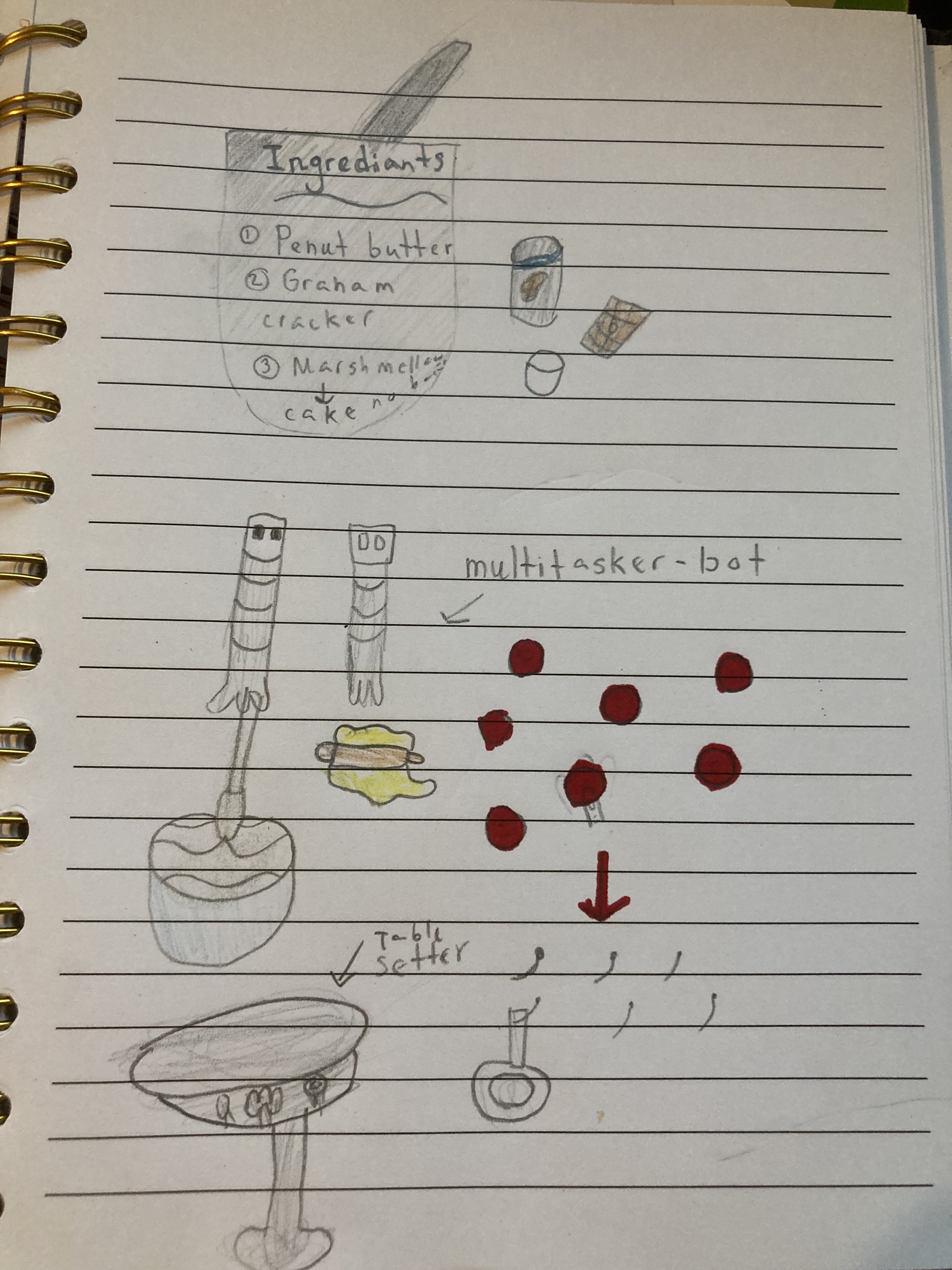}
         \caption{Joanne's (mother from the Booth Family) design sketch, multitasking robotics hand for making crackers}
         \label{fig1b:Joanna's sketch}
     \end{subfigure}
     
        \caption{Sketches showing ways of using technology to automate and reduce labor around the work of feeding}
        \label{fig:}
\end{figure}
As one might predict, many of the women who had shared their struggles with the everyday work and stress of family meals during interviews used the subsequent design sessions to envision technologies that were able to automate all the tedious or time-consuming tasks around meal preparation. Jen, the busy working mother of the Chen family (F9), for instance, sketched a "smart kitchen system" that could assist her with \textit{"all the laborious work"} involved in getting a meal on the table. This labor included routine and mechanical activities like prepping ingredients (Figure \ref{fig1a:Jen's sketch}), as well as the "thought work" necessary for coordinating and planning meals or cooking food.

Describing her smart kitchen system, Jen presented a holistic design vision of both greater efficiency and social support where she could use one interface to view and share recipes with friends and neighbors, generate a grocery list for the meals she wanted to make that week, and place an automated grocery order to deliver any needed ingredients to her doorstep. And if she was too exhausted after a long day to figure out how to put together an evening meal, her smart kitchen could also provide voice-controlled cooking support, allowing Jen to follow a simple audio guide for preparing ingredients step-by-step. And, when it was her husband Steven's turn to make the family meal--there would be no more excuses about him not knowing how to cook--as her kitchen system would have a smart stir-fry pot that could automatically cook her children's favorite dishes! 

Similarly, Joanne, the main meal preparer of the Booth Family (F15), sketched a "multitasker-bot"  with an adaptable robotic hand attachment (shown in her drawing as a whisk) that could complete a sophisticated series of step-by-step cooking movements to bake items like bread or crackers (Figure \ref{fig1b:Joanna's sketch}). While participants like Jen and Joanne welcomed the idea of greater freedom from the responsibilities of domestic labor, they also were wary about how smart technologies might alter their roles as a "good" mother and spouse, which required being able to make good decisions for their family. Of particular concern was how automation might limit a parent's agency and control of family meals. Significantly, in the Chen Family Smart Kitchen system, it was Jen who would continue to closely manage all the health and nutrition-related decision-making for family meals, envisioning a scanning feature that would allow her to read the bar code from any package for nutritional information to keep track of her husband and two sons and \textit{"make sure everyone is eating healthy."} In this way, while there was a strong desire for technology to reduce much of the labor around family meal preparation and planning, particular elements of feeding (even if time-consuming or difficult) were viewed as too deeply enmeshed in familial roles like parenthood to fully automate.

\subsubsection{Reconfiguring Social Interactions}

\begin{figure}
     \centering
     \begin{subfigure}[b]{0.3\textwidth}
         \centering
         \includegraphics[width=\textwidth]{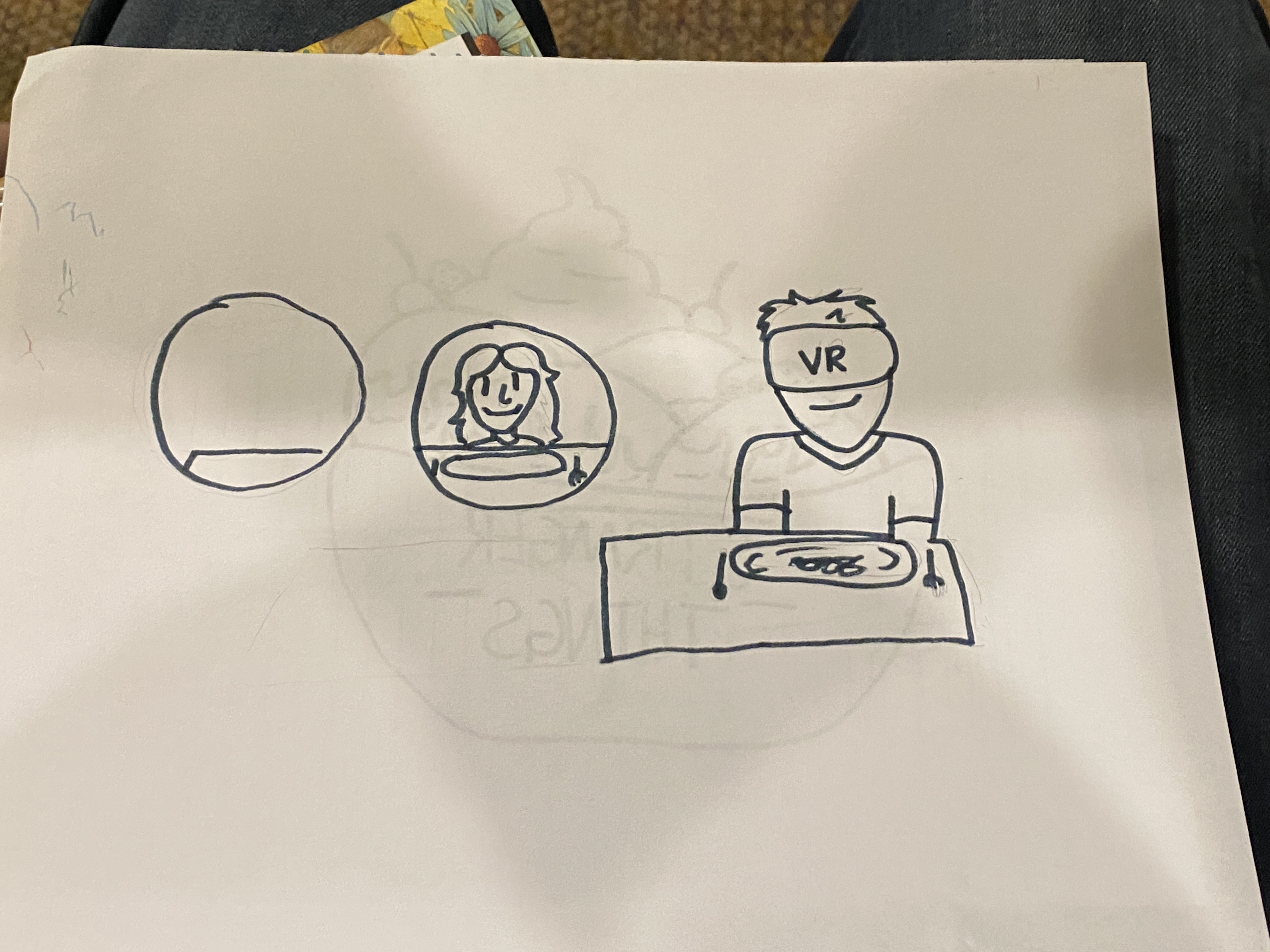}
         \caption{Dave's (father from the Spencer Family) design sketch, eating pills as dinner with a VR headset}
         \label{fig2a:Dave's sketch}
     \end{subfigure}
     \hfill
     \begin{subfigure}[b]{0.3\textwidth}
         \centering
         \includegraphics[width=\textwidth]{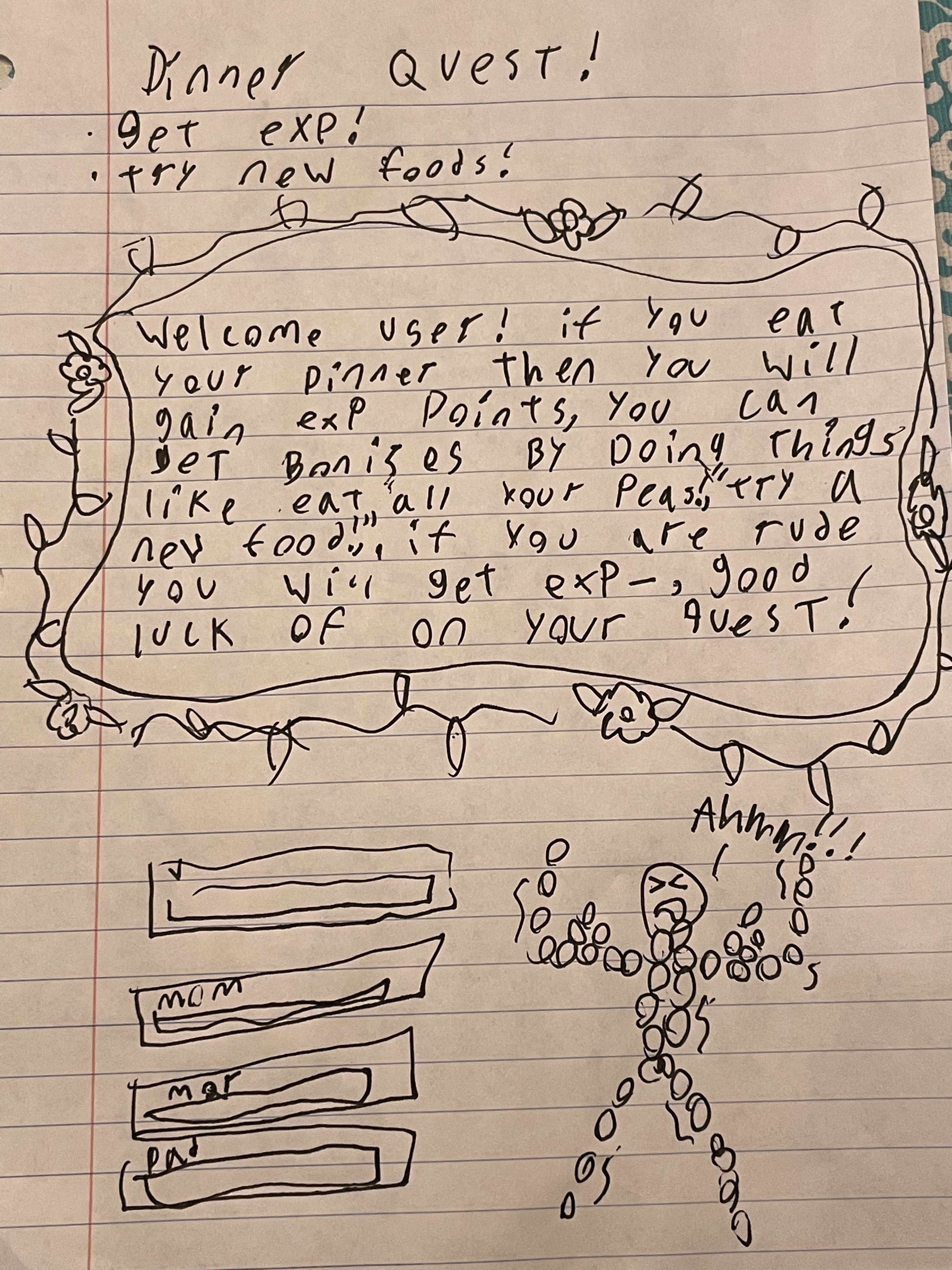}
         \caption{Hannah's (daughter from the White Family) design sketch, a game "Dinner Quest" for eating together}
         \label{fig2b:Hannah's sketch}
     \end{subfigure}
     \hfill
     \begin{subfigure}[b]{0.3\textwidth}
         \centering
         \includegraphics[width=\textwidth]{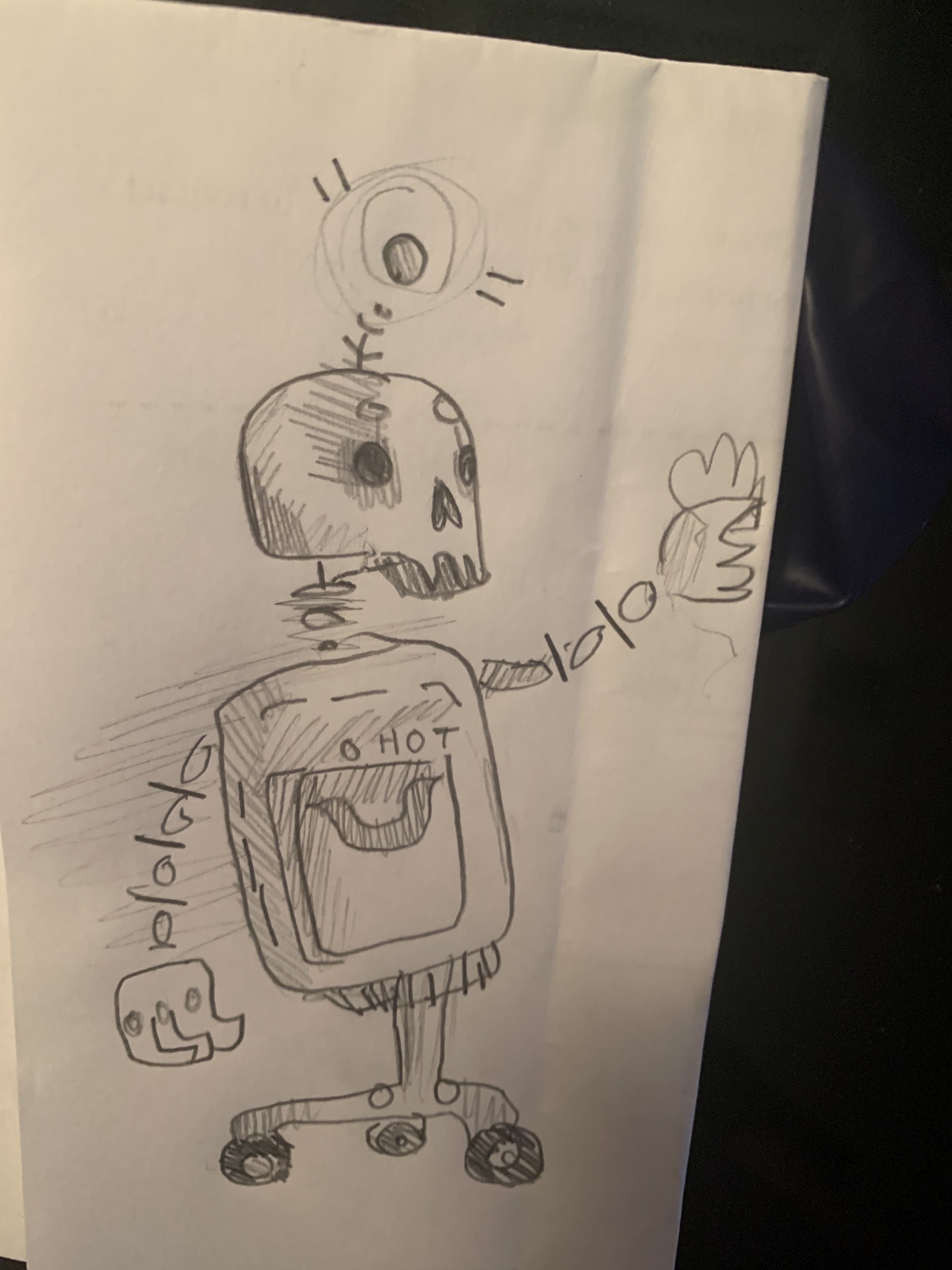}
         \caption{Jess's (son from the Jean Family) design sketch, robot for assistance and companionship in the home}
         \label{fig2c:Jess's sketch}
     \end{subfigure}
        \caption{Different design ideas representing ways to support more meaningful (or fun) social interactions around family meals}
        \label{fig2:Differen design idea to support social interactions}
\end{figure}

Many sketches in design sessions revealed a desire to use technology to support more meaningful (or fun) social interactions with family members, but these design visions were rarely embraced by all members of the family. Indeed, many participants shared deep concerns about how novel meal interactions might disrupt the social and emotional dynamics of the family in troubling ways. For example, Dave, the father of the Spencer family (F2), described feeling like he was missing out on the shared stories and conversations that happened during family dinners when he was out of town traveling for work. In his sketch, then, he imagined using a type of virtual reality (VR) headset to keep him connected to his wife and children during everyday meals. And by having everyone \textit{"eating in the dark"} (e.g., wearing a VR headset during the meal), Dave reasoned, his kids would no longer be distracted by their phones and have more engaging conversations: \textit{"You have to focus on the verbal communication."}  (Figure \ref{fig2a:Dave's sketch}). Dave's partner Jane, however, was distraught with his vision for eating family meals virtually: \textit{"I could never imagine myself doing that. That would be so sad."}  After spending time and effort cooking for her family as the main meal preparer, Jane wanted her children to appreciate her feeding work, including paying attention to the aesthetic experience of the meal (e.g., food texture and flavors) that she worked hard to prepare. Something important was being lost in the family meal if social interactions became disconnected from one's material environment.

We found children primarily imagined technology making the eating experience more playful and engaging. Imagining future family meals, for them, was about exploring ways technology can inject more fun and playfulness into the meal experience. For example, Hannah, a twelve-year-old from the White Family (F5), came up with an interactive system called "Dinner Quest" that gamified the family meal experience (Figure \ref{fig2b:Hannah's sketch}). In her sketches, she mapped out how Dinner Quest would calculate "experience points" for each family member based on their behaviors at the dinner table. An introductory prompt to her system explains the rules: \textit{"Welcome user! If you eat your dinner, then you will gain experience points; you can get (a) bonus by doing things like eating all your peas, try new food. If you are rude, you will get (minus) experience points. Good luck on your quest!"} Hannah thought Dinner Quest would help make dinner more fun as she likes to compete with her parents and sister. Her design incorporated the house rules her parents routinely remind her about (e.g., eat your vegetables, mind your table manners), but that she got tired of hearing. Changing their nagging into \textit{"something more interesting"} inspired her design. There were also many sketches of social robots from kids and adults alike. 10-year-old Jess (F12), for instance, presented a sketch of a family robot (Figure \ref{fig2c:Jess's sketch}) that not only helped out in the kitchen, but acted as a home assistant, companion, and playmate. He described his robot \textit{"as another family member or friend to play with"} and imagined the robot sitting with their family for meals and prompting entertaining conversational topics to "\textit{keep the chat interesting."} These visions corresponded to children's reported desire to make the common parental "check-in" conversations at meals less tiresome. It also reflects their overall marginal involvement in helping with the current family meal work.

Many of these design visions represented uncertainty about how such technology would fit into the everyday rhythms of domestic life, as well as fears about disrupting the social experience of the family meal. Many participants held strong views (and, for some, even a sacred understanding) that family meals were times set aside for emotional bonding and family collaboration. While family meals included moments of fun and entertainment, they also involved teaching children important lessons like table manners (e.g., recognizing people's hard work and appreciating a nice meal). When feeding work is seen as a necessary act of care, however, the social and emotional challenges of the family meal (e.g., boredom, invasive parental "check-ins," even adult disagreements) don't always require a solution.

\subsubsection{Reconfiguring Traditions}

\begin{figure}
     \centering
     \begin{subfigure}[b]{0.48\textwidth}
         \centering
         \includegraphics[width=0.9\textwidth]{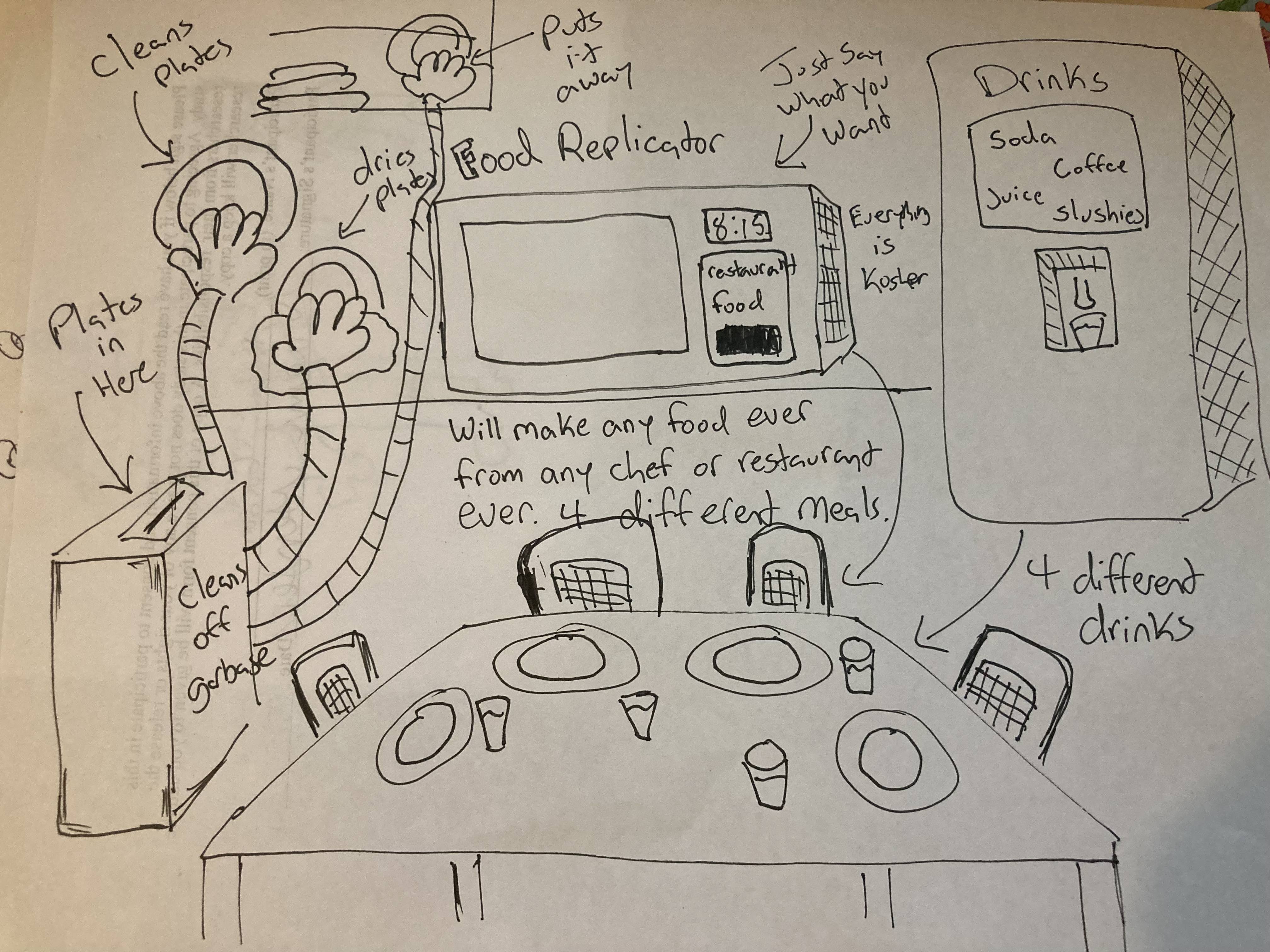}
         \caption{Emma's (mother) design sketch, smart kitchen that handles everything for making a meal}
         \label{fig3a:Emma's sketch}
     \end{subfigure}
     \hfill
     \begin{subfigure}[b]{0.48\textwidth}
         \centering
         \includegraphics[width=0.9\textwidth]{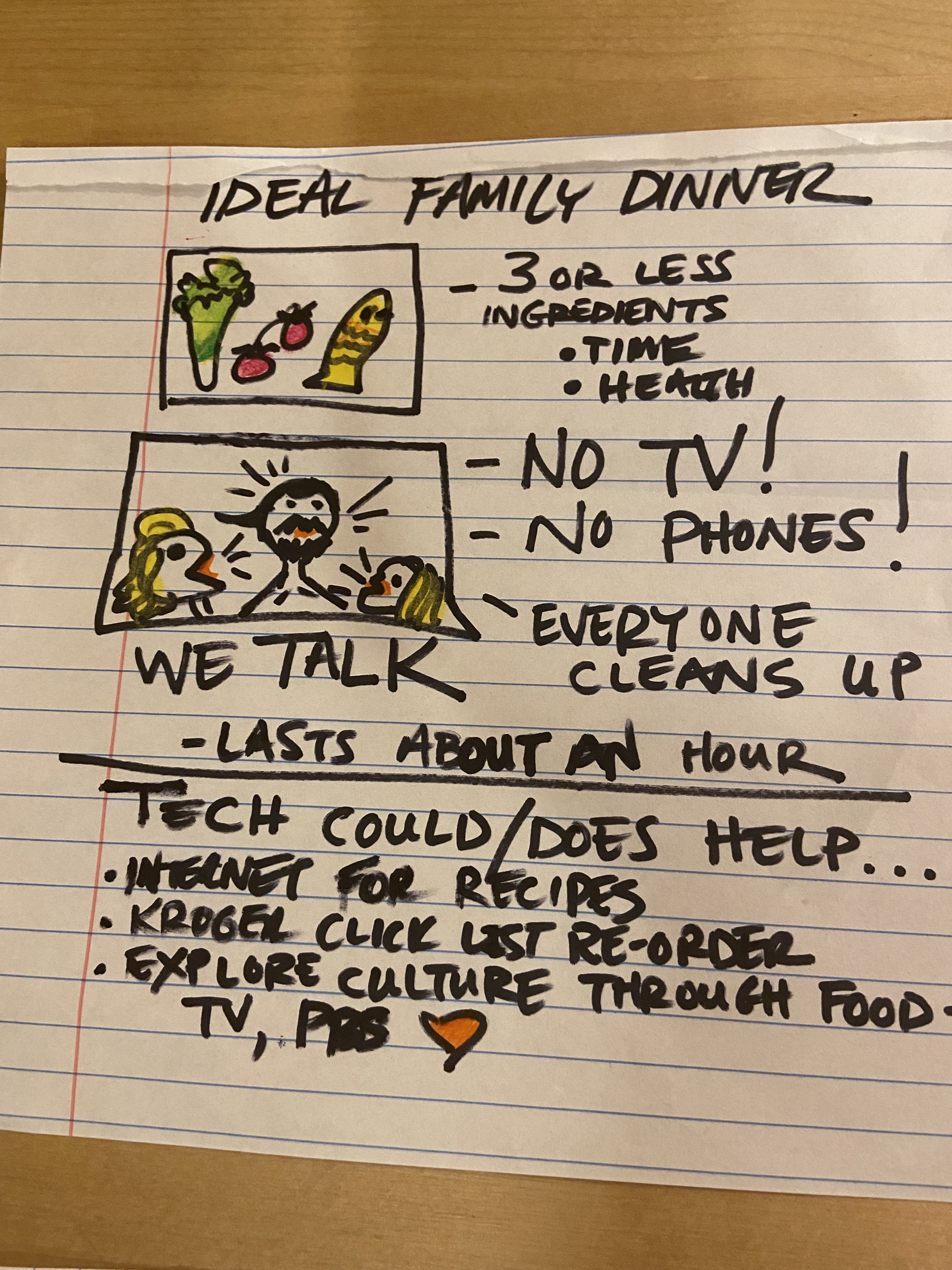}
         \caption{Claude's (father) design sketch, ideal family dinner with limited tech usage}
         \label{fig3b:Claude's sketch}
     \end{subfigure}
        \caption{Sketch from parents from Jean Family showing varying levels of comfort for involving tech in family meals and its potential impact on meal traditions}
        \label{fig3: sketch from Jean family}
\end{figure}

Many families saw meal-related activities not only as an important part of lived experience (e.g., connected to one's heritage or family tradition), but also as meaningful work worth doing in large part due to its very messiness and difficulty (e.g., connected to character-building and cultivating virtues like gratitude and independence). In the Jean family (F12), partners Claude and Emma, who were parents to a 10-year-old child, found themselves wrestling with how technology might change their family's future meal experiences. As the main meal provider, Emma's sketch presented a kitchen system (Figure \ref{fig3a:Emma's sketch}) that epitomizes Weiser's vision for "calm computing" \cite{Weiser1991TheCentury}:

\begin{quote}
    \textit{"Well, imagine this – the system is seamlessly integrated into our existing kitchen appliances and furniture. It can help with basically everything around cooking and stuff...And I don’t want it to be like a typical household robot. I want it to be more invisible, it should act like a robot but it blends into the environment, you know? Like it takes over all the work and just gets the meal ready and serves us. But it’s also not intrusive while it does that. And so mealtime for is is just to relax and enjoy eating together. Nothing else." }
\end{quote}

While Claude agreed that it might be helpful for technology to facilitate some of the routine and simple tasks around meal preparation, like showing recipes or ordering groceries, he found himself particularly concerned about what the "\textit{nothing else}" language in Emma's description would mean for their family relationships. "\textit{We can have the tech to just do all the work for us," }Claude reflected, \textit{"but I don’t know if that’s necessary or ideal." }He explained that for him, there was a lot of value in learning new skills by doing, and he wanted his child, Jess, to likewise have everyday opportunities to develop as a person. His sketch of a family meal included \textit{"working together."}\textit{"It’s about the experience, you know...chopping and mixing things, chatting and sharing about the day, and maybe even teaching} [their child] \textit{a thing or two about the recipe"} (Figure \ref{fig3b:Claude's sketch}). Indeed, Claude feared that the core experience of the family meal--the tradition of cooking together--could easily be \textit{"taken away"} by technology:
\begin{quote}
    \textit{"You gotta learn by doing it yourself…Yeah I think it’s crucial to limit tech use and it definitely shouldn't be omnipotent to a point that it just strips away the core experience of making a meal as a family. [...] I think the process of making a meal is important, it can make for family memories and I think it’s important to keep this tradition[...] I mean, sure, the process of cooking together is messy. It’s almost always guaranteed to be like that. But I think it’s just how it is, you know…struggle is part of the experience."} 
\end{quote}

For Claude, there is a place for technological intervention, but only if it supports (rather than interferes with) the social and emotional complexity of feeding as care work. The celebratory nature of meals (e.g., memories, tradition, and togetherness) is inextricably bound up with messiness (e.g., the everyday discomforts and trials of domestic labor, emotional tensions that come with eating together, and even potential conflicts) that come with family meals by virtue of being a deeply social and collaborative experience. Indeed, in contrast with visions that center on the ease of automation, calmness of seamless integration, or the exciting novelty in human-robot interaction, Claude's ideal family meal is a fraught activity -- the struggle is an essential part of the experience.

In summary, we found that family members all have differing ideas about future meal technology. Participants' visions of meal technologies -- even if the AI-driven or smart-home technology could be designed to seamlessly organize/buy/prepare perfectly nutritious meals -- were rarely (if ever) "solutions" that were acceptable to all members of the family. In fact, many times, these visions resulted in other family members feeling a sense of outrage, dismay, worry, fear, or anxiety over the risks and potential harms such technology may have on the very fabric of family life. In our interviews and design activities, we captured ongoing frustrations with the status quo of family meals (especially with regards to gendered forms of labor), but also a sense that something -- essential to \textit{being a family} -- was at risk of being lost by work obligations, hectic schedules, and the technologies of our present (e.g., cellphones, videogames, and smart devices). The idea of adding more technology to family mealtimes as a way of "fixing" these challenges did not sit easily with our participants (or our research team). In the next section, we take a step back to reflect on the importance of attending to domestic tensions in technology design, both as a means of making useful interventions and creating potential harm.


\section{Discussion}

Meals are often viewed in HCI/CSCW as a celebratory time \cite{Grimes2008CelebratoryHCI, Ferdous2017CelebratoryMealtimes} for families to gather together to eat and socialize -- experiences of commensality that can be positively augmented with technology. While technology design that aims to create positive meal experiences is a valuable goal (indeed, one we share), an overly narrow interpretation of this design narrative can cause many types of meal experiences to be unaccounted for. For instance, many people find themselves feeling routinely stressed, exhausted, angry, or even just bored during family meals, but still continue to view eating together as a central part of domestic life. For every moment of shared togetherness, there are several family meal experiences of frazzled working parents who struggle to conjure up last-minute recipe substitutions, children who seem to find new and creative ways of \textit{not} eating their peas, and parents who feel guilty for wanting to relax with a takeaway instead of cooking a homemade meal for their children. What is the significance of family meals for them if not celebratory? And, how do we, as HCI/CSCW researchers and designers, address (if at all) such familial complexities through technology design?

There is a long scholarly tradition in HCI/CSCW that has argued for the importance of understanding "the social" in system design \cite{Derix2022ItsTime, Moser2016TechnologyMealtimes, Yu2023FamilyComputing, Oduor2014HowKenya}. Paul Dourish and Genevieve Bell \cite{Dourish2011DiviningComputing}, for instance, have written about "embracing mess" in pervasive computing environments like the smart home, noting the stubbornly social practices that cannot be easily designed away (such as varying cultural expectations around privacy, safety, or trust). Their concerns have been picked up and taken seriously among scholars in domestic HCI, particularly those who design for families and children (e.g., \cite{Hiniker2016NotRules, Mazmanian2017OkayAge, Derix2022FamilyRelationships} ). Researchers in this space have all argued for more holistic approaches to design that can account for diverse family settings and needs. We, too, embrace this ethos of "mess" by reflecting on the significance of disagreements and conflicts in family meal experiences. 

We found (unsurprisingly) that meals \hi{with children} are naturally messy situations, in which a wide range of family differences and disagreements emerge. In our study, it was common for family members to hold conflicting views on various aspects of the family meal, including what types of food should be prioritized and valued as healthy, how the family should interact with each other socially, and what forms of labor need to be taken on (and by whom). Furthermore, families also held divergent views on technology as a possible "solution" to these issues. As discussed in section 5.3 (e.g., discussion from the Jean Family, Claude's design sketch), one of our participant-parents discussed his reasons for actively resisting his partner's idea of creating a kitchen system that could automate the tedious tasks of meal-making.  \hi{He explained that he} wanted his children to learn to \hi{how to cook by going through the work of} prepping ingredients and cleaning up dishes \hi{together} as a family. Indeed, his very understanding of \textit{family} might be captured by his reflection: "\textit{Struggle is a part of the experience}." 

Given that struggle can be seen as an inevitable (and at times valuable) part of being in a family, we believe that a critical examination of common domestic tensions in family meals provides a useful starting point for helping to balance and nuance current design approaches for family-food interactions. For example, seeing the activities of feeding as a form of care work helps designers in HCI/CSCW reconsider what might be gained or lost in trying to find technical solutions to the kinds of struggles that make up everyday family life. 

In this section, \hi{then,} we highlight two types of domestic tensions that are a part of everyday feeding work. (We note here that these sensitizing concepts are not intended to \hi{represent an} exhaustive \hi{categorization or list of tensions}, but rather to open up conversation in the wider HCI/CSCW community about \hi{how family meal technologies might be re-conceptualized from a new perspective}). First, we draw on our \hi{study} findings to \hi{suggest possible} design implications around engaging \textbf{\textit{generative discontents}}. \hi{We use this first term to call attention to those} interpersonal differences and disagreements that are a routine and expected part of participating in family meals (e.g., stress around organizing family mealtimes or disagreeing with a partner about eating too much fast food), but which may have social value in struggling through together. Second, we discuss the importance of \textbf{\textit{systemic discontents} }in relation to designing family meal experiences. With this term, we refer to types of harmful or unjust struggles rooted in broader social, political, or historical factors (e.g., inequities in gendered labor or the cost of living crisis which turns family meals into desperate fights to put food on the table). While offering less straightforward opportunities for design intervention, systemic discontents are also costly to ignore. We conclude \hi{our discussion section} with several reflections for HCI/CSCW scholars and researchers to consider in broadening \hi{domestic} design narratives to account for \hi{a wider range of} social and emotional \hi{dimensions in} family food experiences.

\subsection{Generative Discontents: Engaging Family Differences and Disagreements in Design}
Our findings show that everyday disagreements between family members happen across all parts of a meal, from deciding when to serve food, to debating what makes for an enjoyable and meaningful family experience, to negotiating who washes the dishes. As explored through the variety of feeding stories (as described in 5.1), many factors impact how such domestic tensions arise around meal planning and preparation, as well as the particular forms of stress and frustration this brings up within the family. At first glance, such tensions might be considered as "pain points"--those problematic places in the user experience that designers look to resolve (or heal) through technology. We found, however, that tensions are not always \hi{negative experiences}, but can help people articulate important viewpoints, make unseen labor visible, and express emotions. We use the term \textit{generative discontents} then to capture the ambiguous nature and creative possibilities of domestic tensions \hi{in design}.

\subsubsection{Mediating divergent needs, goals, and expectations}
During the course of our study, we often saw family members (like the parents in the Spencer family) learning about each other's \hi{different} perspectives about meals -- often for the first time -- through participating in interviews and design sessions. The struggles they shared (such as differing nutritional goals) are significant to the health and happiness of their family, but \hi{could be understood better as} matters of differing perspectives and personal needs, rather than outright conflicts. Such tensions are amenable to system support. However, our findings also suggest that it is important \hi{for any technological} intervention \hi{to "open up"} space for multiple perspectives to co-exist, rather than offering \hi{a single, top-down} solution to complex family settings. We advocate for a design approach that sees technology as an opportunity to help mediate tensions in family meals by scaffolding the process of sharing and communicating each family member's needs, goals, and expectations on what makes a meaningful family meal experience.

Designers can approach potential tensions and conflicts in family meals in several generative ways. One design implication could be to explore technology's potential for helping facilitate challenging family conversations, allowing households to push past rote parental "check-ins" often disdained by their children, to address difficult or embarrassing topics such as dietary "cheating," feelings of depression or anxiety about home/work life, or address hard issues like concerns about a family member's disordered eating. Such discussions could \hi{easily} feel awkward or uncaring when mediated through an informational pamphlet, or feel hostile and \hi{judgmental} if a family member is uncertain on how best to address sensitive topics. However, socially nuanced and interactive CSCW technologies hold the potential to help mediate some of people's unhappy family moments in thoughtful \hi{and caring} ways. For example, a system centered on generative discontents might support families by visualizing the different feeding tasks and activities each person contributed to the family meal and prompting family members to share how they felt about current meal practices, roles, and responsibilities. \hi{Such a} system might also capture suggestions/changes they would like to propose to existing routines that were frustrating or felt unsupported \cite{wu2023designing}. Family members, \hi{for instance,} might be prompted to reflect on their "discontents" together and have the option to send messages of appreciation/recognition for doing meal-related labor. 

In addition, engaging generative discontents might open up new design opportunities for supporting fun and engaging family interactions in ways that still protect space in the family meal for doing difficult feeding work in the family.
\hi{The observation that children value fun is not particularly surprising, as previous work has discovered children's desire to have fun in learning activities (e.g., \cite{Druga-AI-2022}). However, what is important in our work is when this value of "fun" clashes with other values of family members, such as a parental desire to discuss a failed test at school. How might we design systems to engage with different or even conflicting values and desires in a family? What does it mean to design for the family as a whole when there are multiple preferences to consider? In our conversation with families, we identified opportunities to facilitate fun while also supporting collaborative discussion and decisions within a family.}
For example, expanding on the gamified concept of "Dinner Quest," discussed in our design session (5.3.2), interactive technology can be designed to entertain parents and children through collaborative family meal activities that also allow for family members to \textit{co-create} goals and expectations (e.g., what to eat, when the family meal takes place and for how long). For each such activity, a different person is selected as the “family meal lead” and gets to make the final meal decisions while still having to balance differing goals and needs within the family. In encouraging family members to share their desires and expectations, family meal technologies can become a valuable means of mediating family expectation-setting and role-distribution activities. Importantly, the goal of any type of technology-assisted mediation is not to "eliminate" tensions, but rather, to provide families with support to share their unique views and needs.

\subsubsection{Supporting role flexibility through cooking co-learning with parents and children}
As noted in our findings, mothers overwhelmingly took on the responsibilities of the meal preparer in the majority of families we studied, leaving their partners (which in our study data were fathers) and children with a limited role in meal preparation. While this gendered division of labor was often a result of work schedules (as described in 5.2), we saw in some families, fathers, and older children simply did not know how to cook "as well" as their partner and continued to rely on their skillset (e.g., the Chen Family). As a result, the mothers in our study often mentioned feeling overworked and underappreciated by their families. 

Technologies could prove useful in easing (at least some) of these domestic tensions by supporting the co-learning experience of parents and children who want to learn to cook together. For example, an interactive system might provide families with instruction on how parents and children can collaborate together on making a dish (e.g., the kid washes the tomatoes, and the parent cuts them into thin slices). Different than a traditional cooking tutorial, such a system can add prompts and discussion points throughout the activity, such as fun facts on an ingredient  (e.g., Tomatoes were originally considered poisonous by all but a few Americans until the mid-1800s \cite{FromFarmer}) or open up story-telling opportunities between parents and children (e.g. What was a time when you tried/did something that most people would not do? What was that like and how did you feel about it?). Such features can support parents’ (especially the ones who seldom cook) desire to have more quality time to connect with their kids and also engage young children in interesting activities through family co-learning.  Importantly, such scaffolding features should balance children’s needs/willingness for parents’ support with cooking and parents’ desire for children’s ability to cook independently. As well, co-cooking technologies could spark a teen's interest in the domestic arts (or at least provide them with a greater level of independence through basic cooking skills). How the features can adapt to child development over time or how designers can support negotiating the balance between scaffolding and independence requires more exploration in future HCI/CSCW research.

While we see generative possibilities of discontents for deepening our understanding of family meals as a site of feeding work and for designing technologies that center the social and emotional complexities of care, we also recognize there are many types of discontents that need to be resisted rather than engaged. In the next section, we reflect on a few such tensions present in our study of family meal technologies.

\subsection{Systemic Discontents: Broadening the Design Space of Family-Food Interactions }

Throughout the course of our study, we identified moments in the family meal experience when the tensions or struggles described by participants extended beyond interpersonal disagreements, communication breakdowns, or personal preferences between family members, to include a wider set of socioeconomic and political forces. \hi{Such domestic tensions might be better articulated as \textit{systemic discontents}} to call attention to the wide range of social, cultural, and historical conflicts that shape our domestic environments and impact the intimate relationships between family members, not all of them for the better. We argue that such discontents need careful consideration for those working in the design space of domestic technologies, like family meals.

Our study, for example, found numerous examples of gendered labor around meal preparation and coordination work, most of which negatively impacted female family members. Sometimes problematic views of "women's work" were deeply entwined in people's family histories and cultures, as exemplified by a female participant (Jen's comment in 5.2.1) who spoke about the negative emotional (and physical) impacts patriarchal meal expectations had on the older women in her family. But often, as in the case of the majority of our households, gendered forms of labor remained invisible and unsupported in the home. And when acknowledged, remained unchanged in domestic meal routines. 

While adopting a critical perspective on feeding work in technology design can potentially support greater awareness of domestic labor issues (as explored previously), our findings show that technology (even automation) is not a straightforward solution for the underlying misogynistic attitudes that demean and devalue feeding as "women's work" still found in many homes in the United States, as well as other countries. As many scholars have noted, problematic understandings of gender and labor are often unquestioningly incorporated into the design of new technologies. Science, Technology, and Society (STS) researcher, Thao Phan, for instance, has critically examined the gendered stereotypes of virtual "female" agents like Siri and Alexa, who are programmed to act as personal servants \cite{phan2017materiality}. We see disturbing echoes of the gender biases of Silicon Valley in the sketches our families drew of future kitchen robots that could serve family members any type of food they desired, clean up, and be a playmate and friend. 

\hi{Domestic discontents-- both generative and systemic -- can impact all families. In our work, we saw how a community of Midwestern families in the United States experienced everyday tensions and disagreements around meals. In this vein, one striking implication of this paper is that even the most privileged participants in our study (e.g. those who were part of a dominant social group in the United States or with economic resources such as being upper-middle class, married, straight, or \hi{White}), still \textit{struggled} with family meals.} For most families in our study, various discontents, like the unequal division of labor in the home, were most often a source of daily frustration rather than a deep emotional trauma. For a smaller number of our study participants, however, such as families who had single mothers, or were members of racial minorities, or had nonbinary children, the lived experience of gender, racial, and class inequities had the potential to make their homes a far more precarious places to be. 

Likewise, for families confronted with chronic food insecurities or living with domestic violence, family meals might be a source of despair or even danger amid heightened stresses to provide food for one's family. Making feeding work visible, DeVault argues, involves "a clear and honest assessment of both the value of caring and its darker side--the ways it can diminish the one who does it, and its negative consequences for those who get defined as 'dependent'" \cite{DeVault1991FeedingWork}. An argument that also holds true for how we conceptualize the design of domestic technologies for family-food interaction. \hi{As Star and Strauss argue, making domestic work "visible" through system support might also lead to a wide range of unintended consequences, including increased scrutiny and surveillance for those who must perform it \cite{star1999layers}.}

One of the larger aims of this paper is to deepen our community's understanding of social and emotional complexity in domestic design settings and explore new ways of embracing "the mess" in system design. \hi{\textit{Generative discontents} highlight the design possibilities in embracing tensions, while} the sensitizing concept of \textit{systemic discontents} extends the boundaries of domestic technology settings (i.e., the family home) to account for broader social issues, such as shifting norms around gendered divisions of labor that were discussed here. Further studies in \hi{HCI/}CSCW are needed to determine what discontents are relevant for a wider range of family experiences, including the needs of nontraditional families (such as non-cisgendered, childless, elderly households) that were not represented in our study data. Despite these limitations, we argue that the types of discontents that we draw out in this discussion section offer an intriguing counter-narrative for envisioning new directions in family meal technologies beyond the celebratory visions of commensality \hi{or the idea of corrective technologies that currently dominate this} design space. 

Importantly, \hi{the framing of} discontents reminds us as designers and researchers that not all user perspectives should be reconciled or pain points eased. As in feeding work, embracing tensions is an essential part of the design process \hi{and can be an important act of care on the part of the designer}. \hi{While our study is situated in the context of everyday meal experiences, we would argue that this critical perspective could be useful in more fully conceptualizing family-centered design and other related settings in the field of domestic HCI. Recent work, for instance, has started to explore the various influences that technology adoption could have on family dynamics, and many of those interventions will lead to troublesome tensions in those relationships \cite{Beneteau2020ParentingDynamics, Yu2023FamilyComputing, Liaqat-sharing-dynamics2023}. Our findings highlight the need for HCI/CSCW researchers and designers to better capture nuanced family dynamics for domestic systems that can engage messy communal interactions (e.g. family food fights) and systemic challenges (e.g. gendered labor inequities) rather than idealized food interactions. In this, our work aligns} with other critical HCI/CSCW literature that cautions our community about the risks in seeking \hi{straightforward} technical solutions to thorny human domains \cite{Baumer2011Whentechnology, Dourish2006ImplicationsDesign}. We hope our paper helps designers cultivate a higher tolerance for taking pause and sitting with people's experiences of pain, discomfort, and unhappiness before jumping to overly simplistic or naive forms of support. After all, for many families, struggle is a part of the experience -- a messy reality that can be both celebratory and concerning.

\section{Conclusion}
 In this paper, we offer the HCI/CSCW community a deeper and more nuanced understanding of the context of family meals and considerations for designing domestic technologies that account for social and emotional complexity of this design space. Reporting on empirical findings from interviews and design sessions with 18 families with children (61 participants), our paper focuses on unpacking the differences, disagreements, and conflicts that arose during people's everyday meal experiences, as well as how families imagined technology might be able to address those fraught relations. Drawing on feminist and sociological theories about the care work of feeding \hi{a family}, we conceptualize the domestic tensions found in our data as types of \textit{generative }and\textit{ systemic discontents}. In this, our paper contributes to ongoing HCI/CSCW conversations in domestic technology and family-centered design by offering \hi{novel} design concepts that are useful for critical reflection and creating new types of family meal technologies that center the experiences of those who struggle.
\begin{acks}
We thank Ellie Harrison and Sarah Nikkhah for their help in the early versions of this work. We also wish to express our sincere acknowledgment to the families who shared their personal experiences with us, and whose invaluable contributions enriched this study. This research was supported in part by the National Science Foundation awards IIS-1948286 and IIS-2047432. 
\end{acks}





\bibliographystyle{ACM-Reference-Format}
\bibliography{references.bib, newRef.bib}

\end{document}